\documentclass[12pt,preprint]{aastex}

\begin{document}

\title{Stellar and Circumstellar Properties of Class I Protostars}

\author{L. Prato\altaffilmark{1,2}, K. E. Lockhart\altaffilmark{1,3},
Christopher M. Johns-Krull\altaffilmark{2,3}, John T. Rayner\altaffilmark{2,4}}

\altaffiltext{1}{Lowell Observatory, 1400 West Mars Hill Road, Flagstaff,
AZ 86001; lprato@lowell.edu}
\altaffiltext{2}{Visiting Astronomer at the Infrared Telescope Facility,
which is operated by the University of Hawaii under cooperative agreement
NCC 5-538 with the National Aeronautics and Space Administration, Office of
Space Science, Planetary Astronomy Program.}
\altaffiltext{3}{Department of Physics and Astronomy, Rice University,
MS-108, 6100 Main Street, Houston, TX 77005; k.e.lockhart@gmail.com, cmj@rice.edu}
\altaffiltext{4}{Institute for Astronomy, University of Hawaii, 2680
Woodlawn Drive, Honolulu, HI 96822; rayner@ifa.hawaii.edu}

\begin{abstract}
We present a study of the stellar and circumstellar properties of
Class I sources using low-resolution (R$\sim$1000) near-infrared
$K$- and $L$-band spectroscopy.  We measure prominent spectral lines
and features in 8 objects and use fits to standard star spectra
to determine spectral types, visual extinctions, $K$-band excesses, and water
ice optical depths.  Four of the seven systems studied are close binary
pairs; only one of these systems, Haro 6-10, was angularly resolvable.
For certain stars some properties found in our analysis differ substantially
from published values; we analyze the
origin of these differences. We determine extinction to each source
using three different methods and compare and discuss the resulting values.
One hypothesis that we were
testing, that extinction dominates over the $K$-band excess in
obscuration of the stellar photospheric absorption lines, appears not to be true.
Accretion luminosities and mass accretion rates calculated
for our targets are highly uncertain,
in part the result of our inexact knowledge of extinction.
For the six targets we were able to place on an
H-R diagram, our age estimates, $<$2~Myr, are somewhat younger than
those from comparable studies.  Our results underscore the
value of low-resolution spectroscopy in the study of protostars and
their environments; however, the optimal approach to the study of
Class I sources likely involves a combination of high- and
low-resolution near-infrared, mid-infrared, and millimeter
wavelength observations.  Accurate and precise measurements of 
extinction in Class I protostars will be key to improving our
understanding of these objects.

\end{abstract}


\section{Introduction}

Class I sources represent one of the earliest stages of star formation and
are identified by a rising spectral energy  distribution (SED) at wavelengths
longer than 2~$\mu$m (Lada 1987).  These sources
are deeply embedded within molecular clouds and are very faint or 
undetectable at visible wavelengths because of a thick envelope of 
circumstellar dust.  This material effectively envelops the whole
star and as a result, the light from these young stellar objects (YSOs) is
absorbed and re-radiated in the infrared (IR).  Therefore, most studies 
of Class I objects employ observations in the near-IR or longer wavelength
regimes where these sources are relatively bright \citep[e.g.,][]{gre96,gre02,
dop05,bec07}.

Following the suggestion of Lada (1987), it has been commonly thought that
Class I objects inhabit an earlier evolutionary stage relative to Class
II sources (classical T Tauri stars, or CTTSs).  However, showing that the 
central YSOs in Class I sources display signatures indicative of an earlier 
evolutionary phase compared to CTTSs has not been straightforward.
\citet{ken98} conducted a visible light, low spectral resolution survey of 
Class I YSOs in the Taurus star forming region (SFR).  They identified
preliminary spectral types and luminosities and determined that
for Class I sources these 
properties were similar to those of the Class II objects in
Taurus.  However, these investigators did find a greater frequency and
intensity of outflows as deduced from forbidden line emission in the Class I 
objects.  \citet{whi04} used high spectral resolution data in the visible
to measure the stellar properties of several dozen Class I objects and
reached similar conclusions, although they argue that the larger equivalent 
width values observed in forbidden emission lines from Class I sources might be 
attributable to the effect of circumstellar disk orientation causing obscuration of
the central continuum source \citep{whi07} rather than to a higher
incidence of strong jets.  \citet{whi04} also
found that the veiling and derived
accretion rates were similar for Class I and Class II sources; however,
using high-resolution IR spectra,
\citet{dop05} found that the veiling and associated accretion rates of Class I
and flat-spectrum YSOs are higher than those of the Class II objects.
As a result of these discrepancies, it is unclear how much of the final mass
of a YSO is accreted during the Class I phase.  Episodic events have been 
proposed to account for significant growth over short periods 
\citep[e.g.,][]{ken98}; however, \citet{whi07} argue that the process by
which protostars acquire the majority of their mass is still
unconfirmed.

From an evolutionary point of view, another property that is expected to
differ between the two classes of YSOs is their rotation rates, but
again unambiguous observational evidence for this is lacking.
\citet{whi04} determined that their sample of Class I objects in
Taurus is rotating at $v$sin$i<$35~km s$^{-1}$ on
average, which makes them generally
indistinguishable from Class II sources in this SFR.  However, \citet{cov05}
found that Class I and flat-spectrum objects \citep[apparently transitioning
from Class I to Class II;][]{gre97} in the Taurus and 
Ophiuchus SFRs do rotate more quickly than Class II objects in the same
regions.  To account for rotational slowing between the Class I and Class II 
phases, \citet{mon00} invoked the onset of magnetic disk braking 
\citep[e.g.,][]{kon91,shu94} over the protostellar lifetime.  In the framework
of this paradigm, once disk locking has fully engaged, the star has evolved to
the Class II stage and its rotation has slowed accordingly.  Alternatively,
\citet{mat05} proposed accretion driven winds to account for angular momentum 
changes in young stars.

It seems unlikely that all Class I objects could be Class II
systems with irregular appearances attributable to geometric effects.
Some Class I sources no doubt are simply CTTSs seen through edge on disks, 
but the rest may well be part of an overall evolutionary sequence from Class 
0/I to Class I/II YSOs.  Untangling the evolutionary state and geometric
effects influencing these classifications is, however, apparently
quite complex.

Following the early, low-resolution ($\sim$600) near-IR spectroscopy
studies of \citet[1997]{gre96} on Class I sources there has been a shift to
primarily high spectral resolution  (R$\ga$18,000) studies of 
protostars \citep[e.g.,][]{gre00, gre02, whi04, dop05, cov05, cov06} as
the requisite instrumentation has become available on large
telescopes.  High-resolution spectra permit the measurement of $v$sin$i$,
magnetic fields, and other detailed properties.  Although this work has proved
valuable, the wide wavelength coverage available with low dispersion
spectrographs provides unique leverage in the study of these very embedded
young stars, allowing for the simultaneous determination of the extinction and the
examination and characterization of many spectral features.
For objects in common with high dispersion studies, the results we obtain
here, based on low-resolution IR data, are not always the same, suggesting that
the two approaches are complementary.

While a number of continuum and line emission processes likely play a role
in shaping the emergent spectra of Class I objects, we undertook the current
study to test the hypothesis that extinction is the dominant factor, as indicated
by analysis of the protostar YLW 15A \citep{pra03}. In this current
paper, we determine the properties of a small sample of 8 Class I
YSOs using a number of techniques such as measuring the equivalent width of 
key spectral absorption and emission line features, placing sources in a color-color
diagram using photometric data obtained from the literature, and fitting
our observed spectra to standard star spectra to determine effective
temperature (T$_{eff}$), extinction, and $K$-band excess.  Our goal is to
understand better the properties of protostars and the characteristics
of their surrounding environments using photometry in combination with
low-resolution spectroscopy of atomic and molecular lines, as well as of
solid state features such as water ice absorption at 3.1~$\mu$m \citep{bec07}.
We examine several approaches to measuring extinction in our targets and
compare these with each other and with estimates found in the literature.
In this study we find a wide range in the strength and shape of protostar spectral
features, both emission and absorption lines, from source to source,
even in our relatively small sample of 8 objects.  The principal unifying feature
to the entire set is a rising spectrum across the $K$-band.
The observations and data reduction procedures are described in \S2.  In 
\S3 we present the analysis of the data.  A discussion appears in \S4 and 
the results are summarized in \S5.

\section{Observations and Data Reduction}

Our sample consists of 8 Class I YSOs in the Taurus star-forming region,
two of which comprise a relatively wide binary, Haro 6-10.  Three of the YSOs
were unresolved binaries, IRAS 04239$+$2456,
L1551 IRS 5, and IRAS 04361$+$2547. The objects are listed in
Table 1, along with coordinates and observing dates.  The coordinates reported
here were determined using the Two Micron All Sky Survey
(2MASS) interactive image service.  Initially, using coordinates in SIMBAD, 2MASS
images were found to show a displacement of sometimes 10$''$ or more between
the coordinate center and the stellar image.  By iterating the coordinates until
these matched, we determined the best values for our sample targets.
Haro 6-10 is unresolved in 2MASS images.  The sample was drawn from the
study of \citet{gre96} and selected on the basis of steeply rising $K$-band spectra;
Greene \& Lada identified 7 of the 8 systems as Class Is and IRAS 04489$+$3042
as a flat spectrum source.
In addition, 13 spectral type standard stars were observed and
are listed in Table 2 along with their spectral types,
$K$-band magnitudes, and dates of
observation. These standards were selected to cover a wide range of
low-mass spectral types for use in the model fitting described below.

The data were obtained at the NASA Infrared Telescope Facility on
Mauna Kea in 2000 November 18$-$20 (UT), using SpeX, the facility
low-resolution near-IR spectrograph
\citep{ray03}. A 1024$\times$1024 InSb array was used to record the
spectra. Simultaneous observations of the $K$- and $L$-bands
($\sim$2.0$-$4.2~$\mu$m) were acquired by means of a prism cross-disperser.
A slit width of 0\farcs8 was used for all observations, yielding a
spectral resolution of R $\equiv \lambda$ $/$ $\delta$$\lambda$ = 1000.
The seeing ranged from 0\farcs5 to 0\farcs9. The data were acquired
in pairs, nodded along the slit between frames.
Exposure times varied between 0.51 s and 30 s, and multiple exposures were 
taken in each nod position. Stars of spectral type A0
were also observed for removal of telluric features. Exposures of an
argon lamp were taken for wavelength calibration. Flat field and
dark exposures were also acquired. 

Data were reduced using the REDSPEC
code\footnote{See http://www2.keck.hawaii.edu/inst/nirspec/redspec}.
Normalized flat fields were created by taking the
median of the set of three flat field exposures taken for each star.
The effects of subtracting median dark frames from the flat were
insignificant so this differencing was not performed.  Target exposures
were medianed to create a single image of the spectrum of
each star at each nod position.  For each object
these medianed nod pairs were differenced and divided by the
corresponding flat.  The images were spatially
rectified using fourth-order fits to bright A0 star traces in
each order. When the REDSPEC code could not fit the traces
automatically, as
occurred in an $L$-band order where there was significant
atmospheric contamination, manual fits to the traces
were performed. Wavelength calibration was accomplished by fitting
a second-order polynomial to the positions of identified lines in
the argon lamp line spectrum.  Several rows containing the
stellar spectral data in the rectified, differenced image,
one positive trace and one negative trace, were then summed.
By subtracting the resulting negative spectrum from the positive,
we accomplished a double-difference
procedure that eliminated OH night sky emission line
residuals.  A0 calibrator star spectra, observed at similar
airmasses as the standard and target stars, were used for the
removal of telluric absorption lines. Intrinsic atomic hydrogen lines
were interpolated over in the calibrator star spectrum, which
was then divided into the standard and target star spectra. The
ratio was multiplied by a featureless 9500 K blackbody curve to
restore the original shape of the continuum of the standard or
target star. After reduction, cosmetic improvements (such as
interpolation over bad pixels) were made to the spectra. The
reduced $K$-band spectra of the standard stars appear in Figures 1 and 2.
The reduced $K$- and $L$-band
spectra of the Class I objects appear in Figures 3 and 4.

Four of our target systems are binaries; however, only
Haro 6-10 was angularly resolvable.  When extracting the Haro 6-10~S and N
spectra, the number of extracted rows was
chosen so as to avoid contamination by the companion star.
The spectra of IRAS 04239$+$2436, L1551 IRS 5, and IRAS 04361$+$2547
are composed of the blended light of their primary and secondary components.

\section{Analysis}

\subsection{Target Colors and Corresponding Extinction}

Near-IR magnitudes for the Class I sample (Table 3) were obtained from
the 2MASS point source catalogue \citep{skr06}
and used to construct a J$-$H versus H$-$K
color-color diagram (Figure 5).  Haro 6-10~N lacks a $J$-band
measurement and hence is not shown.  The location of Haro 6-10~S
on the diagram implies strong contamination from circumstellar
scattered light.  All other targets show evidence for a large near-IR
excess and most are highly extinguished.

To measure the extinction from the color-color diagram, we used the
relation $A_{v}=13.83(J-H)_{obs}-8.29(H-K)_{obs}-7.43$, derived in 
\citet{pra03}. This dereddens the targets to the classical T~Tauri
locus, i.e. the intrinsic colors of a young star plus circumstellar
disk system \citep{mey97}, although it does not take into account the
effects of disk geometry (e.g., flaring), disk inclination, or the accretion
rate of disk material onto the star(s).  Also, the classical T~Tauri locus
was determined for K and M stars only and thus is only strictly
applicable to these spectral types.  This process assumes that the envelope
surrounding the Class I system contributes only reddening and that
the intrinsic colors of the star$+$disk system are those of a classical T~Tauri
star.  The extinction results appear in the last column of Table 3 and do
not include Haro 6-10~N because of the missing J band data.

\subsection{Equivalent Widths and Surface Gravity}

Equivalent widths for spectral features in the $K$-band were measured
for the standard stars and for the Class I YSOs and are given in Tables 4
and 5, respectively.  The equivalent width of the Br$\alpha$ emission line 
for the Class I sources is also given in Table 5.  Equivalent width 
uncertainties were determined by
varying the location at which the relative continuum was measured.  For the
$K$-band, this yielded uncertainties of $\sim$0.5 \AA~and for the $L$-band, 
$\sim$2 \AA.  When no value is given in Tables 4 or 5, the equivalent width 
is less than the uncertainty or the line is not detected at all.

Only two targets in our sample, L1551 IRS 5 and IRAS 04489$+$3042,
showed NaI, CaI, and CO (2$-$0) all in absorption.
We compared the locations of these two stars with the dwarf (from our data)
and giant \citep[from][]{wal97} star NaI$+$CaI versus CO (2$-$0)
equivalent width loci (similar to the loci data in Prato et al. 2003)
and find that the Class I sources are 
consistent with giant star surface gravities.  We conclude that,
for Class I protostars, at least in some cases, low surface gravity standards
would be preferable as spectroscopic templates.

\subsection{Spectral Types, $K$-band Excess, and Extinction Revisited}

To determine the underlying spectral type, $K$-band excess, $r_K$, and
extinction down to the protostellar photosphere, we compared the $K$-band
YSO spectra to a suite of dwarf, subgiant, and giant spectral type standards
(Figures 1 and 2).  We initially followed the same procedure described in
\citet{pra03} which we briefly review.  For a range of trial $A_V$ and $r_K$
values, the Class I target spectra are dereddened using the reddening
law of Rieke and Lebofsky (1985) and the veiling 
continuum is removed.  This modified spectrum is then fit to spectral type
standard templates by determining the multiplicative scale factor that
minimizes $\chi^2$.  The combination that gives the lowest overall value of
$\chi^2$ then gives us an estimate of $A_V$, $r_K$, and the spectral
type approptiate for each Class I source.  This procedure provided initial 
estimates of these key properties; however, additional experimentation 
with wavelength dependent veiling and visual examination of the resulting 
fits was used to arrive at the final values shown in Table 6.
Crude uncertainties on the order of 5$-$15\% were estimated from
visual inspection of the YSO fits to the comparison standard stars. Some
comments on the fitting process for each target follow. 

\subsubsection{IRAS 04016$+$2610}

\citet{whi04} and \citet{dop05} identify a K star spectral
type and \citet{dop05} find a large $K$-band excess
for this protostar.  Given the
similar depths of the CaI, MgI, and first CO bandhead evident in the
spectrum of IRAS 04016$+$2610, an early K spectral type appears
to provide the best match.  The $K$-band excess has a positive slope
of $F_{\lambda}/\lambda\sim$0.5 and a y-intercept of 0.1.  Unfortunately,
the subtraction of such a large $K$-band excess from the observed spectrum
results in a noisier spectrum that the original; higher signal to noise
data would improve the accuracy of the fit.  Experimenting with multiple
spectral type standard spectra, a range of extinctions from 43$-$53 mag, and 
veiling at 2.2~$\mu$m of 0.5$-$1.5, we conclude that IRAS 04016$+$2610
is best fit with a K3 spectral type, $A_v=45$ mag, and $r_k=1.3$.
This spectral type and excess are consistent with the results of
\citet{dop05} using high-resolution (R$=$18,000) $K$-band spectra.

\subsubsection{IRAS 04181$+$2654A}

The data of \citet{bec07} taken of this source $\sim$3 years after
our spectra, and at higher signal to noise, show better-defined
stellar absorption features.  However, both the shape of the continuum
and the depth of absorption lines, shown in Figure 3, provide
important limits from which to determine
the best fit.  The presence of water vapor in the atmospheres of M type 
objects produces a break in the slope of the spectrum at about 2.3~$\mu$m 
\citep{wil99}.  Thus, objects with such a break (Figures 1 and 3) are unambiguously of 
later spectral type.  In lieu of clearly detectable stellar absorption
lines, we estimated an upper bound for the extinction by assuming
a zero $K$-band excess and fitting the slope of the continuum to an
M star.  The continuum spectrum of IRAS 04181$+$2654A is consistent with an
early M star and an $A_v$ upper limit of 34 mag.  \citet{bec07} found an
$A_v$ of 18 mag and an M3$^{+2}_{-1}$ spectral type.

\subsubsection{IRAS 04239$+$2436}

All spectral features, including Na~I and CO, observed in the $K$ band for this
target are in emission.  No break in the continuum is detected, thus
a spectral type later than K7 (Figure 1) is unlikely.
Using a K0 V standard and setting the $K$-band excess
to zero we estimate an upper limit for the visual extinction of $\sim$45 mag.

\subsubsection{Haro 6-10~S}

We found a constant $K$-band excess of $\sim$0.5 for Haro 6-10~S and
an extinction of 30 mag.
No break in the continuum was observed, implying a spectral type 
earlier than M0.  Small absorption lines of NaI, CaI, and MgI are
clearly visible; however, the CO is probably in emission,
filling in what would otherwise be fairly deep absorption
features \citep[see][]{dop08}.  We assign Haro 6-10~S an early K
spectral type and use a K3 type to estimate an approximate temperature,
although this is highly uncertain.

\subsubsection{Haro 6-10~N}

Setting the $K$-band excess to zero we estimated an
upper limit for the extinction and the
spectral type of the emission line object Haro 6-10~N.  We find
an upper limit of 59 mag for the extinction and a spectral type
consistent with an early K or G type star.

\subsubsection{L1551 IRS 5}

Figure 6 shows our fit of L1551 IRS 5 to an M3 giant standard
after acccounting for an extinction of 28 magnitudes and a
constant excess of 1.0 across the $K$-band.

\subsubsection{IRAS 04361$-$2547}

With a $K$-band excess of zero we find an upper limit of 30 magnitudes
for the extinction to IRAS 04361$-$2547; the curvature in the spectral
continuum implies a spectral type later than M3.

\subsubsection{IRAS 04489$+$3042}

A hybrid standard star spectrum obtained by averaging the M3III and
M4V spectral type standards provided the best fit to this source
after accounting for a constant $K$-band excess of 1.4 and 18 magnitudes
of extinction.  The deep CO bandheads and characteristic M star
break at $\sim$2.28~$\mu$m support this conclusion.  We find a
{\it negative} slope of 0.5 for the excess across the $K$ band and
a y-intercept of 2.5.

\subsection{Ice Band Absorption and a Third Extinction Estimate}

The most obvious feature in the $L$-band spectra of Class I YSOs is the wide,
deep, water ice absorption at 3.1~$\mu$m (Figure 4).
This feature appears in all 8 of the target sources to varying degrees.
In order to use the 3~$\mu$m ice feature to estimate the optical
depth through the envelope, it is necessary to isolate the ice
absorption profile from the overall shape of the spectral
energy distribution (SED) in the 2 -- 4
$\mu$m range.  To do this, we construct a simple model which is fit
to each source.  The model consists of a template standard star to which we
add continuuum veiling; this sum is then subjected to reddening.
The reddened spectrum is then scaled by a multiplicative factor to
account for the distance to and size of each source.  The template star
spectrum is our observed spectrum of either HR 753 (K3V) or GJ 876 (M4V), 
depending on which is closer in spectral type to our estimate for each 
Class I source.  The reddening law is again that of \citet{rie85}.
The veiling emission is taken to be quadratic in wavelength.  We then
use the technique of Marquardt \citep{bev92} to
determine the best fitting parameters which match our model spectrum 
to the spectrum of each Class I source.  In the fitting procedure, we
fit only the spectral regions shortward of 2.5~$\mu$m and between
3.65 and 4.1~$\mu$m.  For Haro 6-10~N, we also ignore two strong
absorption features, likely artifacts of the reduction, which appear
at 3.76 and 3.88~$\mu$m.  Once we
have fit each observed spectrum, we divide that spectrum by the
fit and set the minimum value of the ratio equal to $e^{-\tau_{ice}}$ and
solve for $\tau_{ice}$.  An example of our fit to IRAS 04016$+$2610
appears in Figure 7.

\citet{whi88} found a linear correlation between $\tau_{ice}$
and the visual extinction, $A_{v}$, for the interstellar medium.
\citet{sat90} examined this correlation for a sample of protostars,
assuming the ice to be in the circumstellar material surrounding the
star itself, and used the same relation as \citet{whi88},
$\tau_{ice} = 0.093(A_{V} - A_{Vc})$, adopting a critical visual extinction
of $A_{Vc}=1.6$ mag.  We use this modified relation to calculate the
extinction as traced by the circumstellar ice feature.  The results
appear in Table 7.

The combination of near-simultaneous moderate resolution
spectroscopy from the near to the mid-IR allows for
the comparison of solid state features and hence reveals 
physical and compositional properties of Class I envelopes
\citep[e.g.,][]{fur08}.  We examined the correlation of the
10~$\mu$m silicate absorption feature strength, $F_{\lambda_1}$
from \citet{kes05}, calculated from a Gaussian fit to the
normalized mid-IR spectra, with
optical depths we determined for the 3.1~$\mu$m
water ice feature (Table 7).  Figure 8
shows a trend towards higher ice optical depths for strong 
silicate absorption sources only; no significant correlation
is obvious.

\subsection{H-R Diagram}

To compare the relative masses and ages of our Class I sources
with those of Class II objects, we plotted stars for which we have
the most reliable spectral types, determined from both absorption lines
and continuum shape, on the H-R diagram
(Figure 9).  For four targets we used luminosity estimates from
\citet{fur08}.  For two others, IRAS 04016$+$2610 and Haro 6-10~S, we
used values from \citet{dop05} either because the value from
\citet{fur08} was anomalously low (implying a main sequence age
for IRAS 04016$+$2610) or because \citet{fur08} did not observe
the source (Haro 6-10~S).  We used our spectral type estimates
(Table 6) combined with the dwarf conversion (Table 2) given
in \citet{joh66}, to determine T$_{eff}$.  Interestingly, Johnson's spectral type to
T$_{eff}$ conversions closely parallel those derived for young
stars by \citet{luh03}, from G through mid-M spectral types.
In Table 8 we list our T$_{eff}$, and for comparison the
luminosity of each star given by \citet{fur08}, \citet{dop05}, and/or
\citet{whi04}.  The resulting mass and age from the pre-main
sequence tracks of \citet{pal99} are provided
in Table 8 for the six targets for which these data were possible
to derive.

Of the seven systems studied in this paper (taking Haro 6-10~S and N as
one system), four are close binaries (Table 1).  The three single systems
in our sample along with Haro 6-10~S and two of the unresolved 
binaries appear in Figure 9.  There is an inherent
overestimate of the stellar luminosity for the unresolved pairs, which are
treated as single objects by Furlan et al. (2007).  Also, geometrical effects
produced by the relative orientations of the circumstellar disks of 
each binary component could lead to anomalous measurements of
extinction and excesses, for example, in the case of a configuration in
which one star is obscured by the disk of the other.  To the 
extent that circumstellar disks (flared or seen nearly edge on) contribute
to the extinction, it is quite possible the extinction is very different to
the two members of the binary.  Without angularly resolved
observations of the component objects in these systems, the degree to
which binarity distorts the results of Class I studies, specifically the
effective temperature and luminosity estimates and thus a source's location
in the H-R diagram, is unknown.  The unresolved binaries are located far above the 
evolutionary model tracks in Figure 9, suggesting unrealistically young
ages for these systems.

\subsection{Accretion Luminosity and Mass Accretion Rate}

For comparison with other studies of Class I objects as well as with Class II
targets, we calculated the accretion luminosity, $L_{acc}$, of the eight 
stars in our sample using the correlation between $L_{acc}$ and Br$\gamma$
line luminosity found by Muzerolle et al. (1998).  Our calculation is
based on the $K$-band magnitude, the Br$\gamma$ emission line
strength, and the overall extinction, for which we used the value determined
by model fitting to standard stars as described in \S 3.3.  The results
appear in column 2 of Table 9.  In most cases, the measurement
is significant only at the one sigma level because the inherent uncertainties
in the relationship between the accretion luminosity
and the Br$\gamma$ line luminosity \citep{muz98} are large.
For half our sample we use upper limits for the extinction; if instead we use the
values for extinction determined from the ice feature (Table 7) we obtain
lower values for $L_{acc}$ in the cases of IRAS 04181$+$2654A
(log($L_{acc}$)$=-$0.84$\pm$1.12), IRAS 04239$+$2436
(log($L_{acc}$)$=-$0.19$\pm$1.06), and Haro 6-10~N
(log($L_{acc}$)$=-$0.94$\pm$1.14).  For IRAS 04361$-$2547, the other
system with an upper limit only for $A_v$ in Table 6, the ice feature $A_v$
(24.2 mag) is similar to the upper limit ($<$30 mag).

For the six targets in our sample for which we have estimates of both
T$_{eff}$ and the luminosity \citep{fur08}, we calculated the stellar
radius, R$_*$, and hence the mass accretion rate, \.M,
following \citet{gul98}: $\dot{M}=1.25L_{acc}R_*/(GM_*)$
(Table 9).   Stellar mass (M$_*$) was estimated from
the location of the targets in the H-R diagram.  Primarily as a result
of the very large uncertainties in the calculated accretion
luminosity, L$_{acc}$, uncertainties in our mass accretion rates
are all approximately an order of magnitude.

\section{Discussion}

\subsection{Comparison with Previous Results from Literature}

\subsubsection{Extinction}

As demonstrated here and also
shown in \citet{bec07}, different approaches to measuring the extinction
along the line of sight to Class I protostellar photospheres produce widely
different results.  We have used the position of sources in a near-IR 
color-color diagram,
fitting to absorption features and the continuum shape of the $K$-band
spectra, and fitting to the $L$-band water ice feature to determine $A_v$
and have found values that differ significantly.  Figure 10 shows the
extinctions determined from de-reddening objects plotted on the 
color-color diagram and from $K$-band spectral fitting as a function of $A_v$
determined from fitting the $L$-band water ice feature (Figure 7).
Color-color diagram determined extinctions appear to scatter
randomly, whereas extinctions
measured from our $K$-band spectra, excluding the upper limits, show some
positive correlation with the ice $A_v$.  This suggests that the color-color
diagram approach does not trace the same material along the line of sight.
The color-color approach was based on the possibly flawed premise that the
circumstellar envelope present in a Class I system contributes only reddening
and assumes that the intrinsic colors of Class I protostars are the same as those of
classical T~Tauri stars.  As a counterexample, consider objects that fall below the
classical T~Tauri star locus in the color-color
diagram, i.e. IRAS 04016$+$2610 and Haro 6-10~S.  These are no doubt
extinguished; however, the effect of scattered light in the surrounding
circumstellar environment  yields unusual colors.
Measurements of the extinction from the optical depth of the water ice feature
in the $L$-band probes the extinction through the region where this
feature forms, while possibly missing extinction caused by dust in inner, warmer
regions of an envelope or disk where it is too hot for water ice to survive.
This could explain the systematically higher $A_v$
values from the $K$-band spectroscopic
fit approach (again, excluding the upper limits) which show a similar slope
as the one-to-one correlation and presumably take into account all
sources of extinction along the entire line of sight to the target system.

Comparison of our extinction results with those of \citet{whi04} and
\citet{bec07} reveals that these studies are generally consistent
with our smaller estimates of extinction based on
fitting of the $L$-band ice absorption 
feature.  \citet{whi04} provide extinctions determined from J$-$H colors
and \citet{bec07} derives extinctions using several different approaches, 
including a spectral fitting procedure similar to the $K$-band fits
employed here.   In their \S 3.8, \citet{dop05} describe an approach
to estimating extinction, based on $K$-band magnitudes, in order to
derive stellar luminosities; however, they do not provide values of
extinction in their paper.  It is difficult to know which approach yields 
the ``correct'' extinction to a protostar.  Ideally, a combination of 
near-simultaneous high- and low-resolution spectroscopy could provide
careful measurements of underlying absorption line ratios as well as a
measurement of the continuum slope.  Even relatively small
uncertainties in extinction can have a strong impact on derived Class I
object properties, particularly accretion rates,
and hence studies of these targets can be significantly
improved with more attention to this problem.

\subsubsection{Spectral Types, Surface Gravity, and $K$-band Excesses}

Three of our targets, IRAS 04016$+$2610, Haro 6-10~S, and IRAS 04489$+$3042,
for which we determined T$_{eff}$, $A_v$, and $r_k$ were also analyzed by
\citet{whi04}, \citet{dop05}, \citet{luh06}, and \citet{dop08}.  Another, L1551 IRS 5,
was analyzed by \citet{dop05}.  Our results for spectral types,
surface gravities, and $K$-band excesses typically agree to within
1$-$2$\sigma$ with those from the literature with the exception of Haro 6-10~S and
L1551 IRS 5, discussed below.  For IRAS 04489$+$3042, \citet{luh06} found
a spectral type of M3.5$-$M4.5 on the basis of visible light observations,
and M3$-$M4 on the basis of IR observations, in excellent agreement
with our determination of an M3.5 spectral type for this source.

For Haro 6-10~S, \citet{dop08} used
high-resolution, high-signal to noise $K$-band spectra to derive a
T$_{eff}$ of 3800~K, cooler than our value of $\sim$4800~K, a
surface gravity (log~$g$) of 4.0, and a $K$-band excess of 2.5,
substantially larger than our uncertain estimate of 0.5.
Based on the appearance of their spectra, these values
for Haro 6-10~S are more reliable than our findings and result in
a mass estimate smaller than ours by a factor of $>$3; both studies
result in similar ages (2$-$3 Myr).

For L1551 IRS 5, we find an excellent match to an M3 III
spectral type.
\citet{dop05} find a T$_{eff}$ of 4800 K, corresponding to an early K
spectral type, and a log~$g$ of 4.0.  In both our analysis
and theirs, the $K$-band
veiling is $\sim$1.  Figure 6 shows our M3 III fit of the modified young star
spectrum, with $A_v=$28 mag and a constant $K$-band excess taken into account.
Inspection of the spectra presented in \citet{dop05} shows an inconsistency
with their earlier spectral type and higher gravity, possibly as the result
of a degeneracy between the equivalent widths of the Na I and Mg I/Al I lines
as a function of T$_{eff}$ and log~$g$ (Doppmann 2007, priv. comm.).  
The lower surface gravity found in
our fit is also consistent with our comparison of the location of L1551 IRS 5
on a plot of the Na I $+$ Ca I versus CO (2-0) equivalent widths (\S 3.2):
it clearly lies along the giant star locus.

In general, from our $K$-band spectral fits we find higher extinctions and
lower excesses than those derived in previous Class I studies. 
We have used arbitrary slopes for fitting the $K$-band excess, with no
physical basis in a realistic model of circumstellar disk and/or envelope
radiation.  A combination of high signal-to-noise ratio low- and
high-resolution spectra and detailed modeling of the
spectral energy distribution, as in \citet{fur08}, will be
necessary to improve these uncertain results.

\subsubsection{Accretion Luminosities and Mass Accretion Rates}

For six of our targets, published accretion luminosities are 
available for comparison in \citet{muz98} and \citet{bec07};
these are provided in Table 9.  Although the procedure for measuring the
mass accretion rates followed in this paper and in the work of 
\citet{muz98} and \citet{bec07} was the same, i.e. based on the Br$\gamma$
emission line, our values are all larger by $>$1$-$2~$\sigma$.
Four objects have upper limits only for the spectroscopically determined
$A_v$.  We recalculated the accretion luminosities for the three largest
upper limits using the $A_v$ values based on the 3.1~$\mu$m ice feature (see \S 3.6).
The recalculated $L_{acc}$ values are much closer to the values found by
\citet{muz98} and \citet{bec07} for the same targets,
illustrating that the relatively large values and upper limits for
the extinction determined from our $K$-band spectra are responsible for our
relatively large accretion luminosities (Table 6).  The impact of these larger 
accretion luminosities propagates into our calculation of the mass accretion
rates, which range from $\sim5\times10^{-7}$ M$_\odot$ yr$^{-1}$ to 
$\sim2\times10^{-8}$ M$_\odot$ yr$^{-1}$ (Table 9), calculable for six
of our targets.  For the targets with published accretion rates,
our estimates are equivalent or larger by up to two orders of
magnitude (Table 9), however, with an
uncertainty of approximately an order of magnitude.  Given this large
uncertainty, our small sample size, and the ambiguity in the
determination of $A_v$, it is impossible to draw
definitive conclusions regarding potential enhanced accretion in
Class I versus Class II objects.  Class II mass accretion rates are
typically $>$10$^{-8}$ \citep{her08}.  Our results point to larger
accretion luminosities and mass accretion rates than those
of Class II objects, suggesting an evolutionary progression
from protostars to classical T~Tauri stars, as discussed in \citet{dop05}
for example, but without the signal to noise to support this
conclusion at a significant level.

\subsection{Accreting Protostars or Misclassified T~Tauris?}

Recent studies of Class I protostars differ in their conclusions regarding the
nature of these objects. \citet{whi04} suggest that most Class Is
have moved beyond their main accretion phase and might actually be Class 
II stars seen edge on through circumstellar accretion disks. 
\citet{dop05} conclude that Class I objects are indeed actively
accreting protostars, albeit spanning a range of accretion activity.
Since the publication of both papers, their authors have
reached some consensus and conclude in \citet{whi07} that,
at least for the Taurus SFR, one-third to one-half
of the Class I objects are likely to be misclassified Class II stars
seen through optically thick disks.
\citet{whi07} also conclude that most Class I sources posses
disk accretion rates below the expected envelope infall rates.
Misclassified T~Tauri stars would certainly show low disk accretion rates;
however, such objects would typically be sub-luminous by a factor
of five compared to bona fide Class Is \citep{whi03}.

Another possibility raised in \citet{whi07} is that our knowledge
of disk accretion rates and mass infall rates is incomplete.
The uncertainties in these calculations are significant:  in this
paper as well as in the literature these values are only known
to a precision of about an order of magnitude.  Our finding of
relatively large accretion rates (Table 9) compared to previously derived
values for Class I objects is not significant.
Objects that are found to be highly extinguished, such as our results
suggest, naturally have large accretion luminosities.
However, the large uncertainties associated
with the mass accretion rate estimates mean that we cannot be
confident that 
our results indicate that these protostars are currently experiencing
a particularly active accretion phase.  Our targets do, however, lie
relatively far up on the evolutionary tracks in the H-R diagram
(Figure 9), with ages $<$2~Myr, consistent with a population younger
than that of the Class II T~Tauri stars with ages of $\sim$1 to
several Myr.

Another source of uncertainty in the determination of Class I
protostar properties is their variability.
Class I YSOs show considerable variation in their spectra, fluxes,
colors, and excess IR emission.  \citet{lei01} studied
the variability of the Haro~6-10 system over a  twelve year period
and found substantial change in the magnitudes and colors of the
two components, especially at shorter wavelengths.  $K$-band veiling
measurements for the Class I protostar YLW 15A (IRS 43) varied
significantly over $\sim$5 years, possibly on time scales as
short as weeks \citep{luh99,gre02,pra03}.  These results suggest
that the geometry of the obscuring material in front of the protostellar 
surface may be shifting rapidly, perhaps producing variations in the
extinction along the line of sight to the protostar
\citep[e.g.,][]{lei01,bec01}.

Consistent with a younger evolutionary stage for the Class I
sources is the ubiquitous atomic and molecular hydrogen emission
in all but one of our targets (Figure 3 and Doppmann et al. 2005).
The atomic hydrogen emission is likely associated with disk accretion
onto the central star and the molecular emission with shocks created
by infalling or outflowing gas interacting with the
circumstellar environment.  IRAS 04489$+$3042 shows only a small
Br$\gamma$ emission line and does not reveal any molecular
hydrogen emission, either in our study or in that of \citet{dop05}.
All other targets have the H$_2$ $v=1-0$ S(1) line in emission and
about half also show emission in the H$_2$ $v=1-0$ S(0) line.

\section{Summary}

We obtained $K$- and $L$-band spectra of a sample of eight
Class I sources in the Taurus SFR with the
SpeX spectrometer on the NASA IRTF 3 m telescope.
We measured absorption and emission
line equivalent widths, determined the optical depth of the 
$L$-band water ice feature, and, where possible, estimated the spectral
type, visual extinction, and $K$-band excesses for each of our targets.
Our analysis and results are summarized below.

-- Using our derived spectral types in combination with 2MASS photometry 
and luminosity estimates available
in the literature, we placed our targets in the H-R diagram and used
evolutionary tracks of \citet{pal99} to estimate stellar masses and
ages.  In addition, we calculated accretion luminosities from the 
Br$\gamma$ line luminosity and determined the associated mass accretion
rate for our targets.  The six objects which we were able
to place on the H-R diagram show ages of $<$2 Myr and clump into two
groups with masses $\sim$0.2~M$_{\odot}$ and 1.9~M$_{\odot}$.

-- For two objects we obtain significantly
different results from those presented
in the literature, Haro 6-10~S and L1551 IRS 5.  The majority of our
targets are multiples; although we resolved Haro 6-10~S and N, 
\citet{dop08} suggest that the southern component is a spectroscopic
binary.  The high-resolution, very high signal to noise data of
\citet{dop08} indicate a later spectral type, closer to M0, than the
early K type found by us.  The companion to L1551 IRS 5 is seen only
at very long wavelengths \citep[$>$submillimeter;][]{loo97} and it is thus unlikely
that this component contaminates the primary spectrum in our
analysis.  However, we found a robust fit for this target to
an M3III spectral type standard, a significantly lower gravity and
later spectral type than the $\sim$K3V type of \citet{dop05}.

-- Our experimental hypothesis, that obscuration of the photospheric 
absorption lines in Class I targets is dominated by
extinction local to the sources, is not correct.  Veiling of lines from
the near-IR excess also plays a key role.  It is likely that emission,
for example from warm CO in a circumstellar disk, also
fills in spectral absorption lines in some cases, hampering the determination
of the underlying protostellar characteristics.

-- We compared our derived properties with values from the literature
and find that our larger estimates of extinction lead to larger accretion
luminosities and larger mass accretion rates.  {\it We stress that the
very large uncertainties, inherent in the relationship between
emission line luminosities and accretion luminosities \citep{muz98}
as well as in our observations,
render these estimates highly uncertain, not only in our study but
also in numerous examples in the literature.}

-- The higher mass accretion rates we find are more consistent with the
interpretation that Class I protostars are undergoing a relatively active
mass accretion phase; however, because of the large
uncertainties in these results,
additional observations are required to confirm this interpretation.
High-resolution, high signal to noise
IR spectroscopy provides a promising approach, however,
low-resolution data across the near-IR may be particularly useful for the
evaluation of extinction.

-- The visual extinction to protostars appears to be an extremely
important and poorly determined property, impacting estimates
of all the key stellar properties including luminosity, mass, age, 
accretion luminosity, and mass accretion rate.  It
therefore represents a critical area for improvement.

\vspace{12 pt}

We thank G. Doppmann, E. Furlan, E. Gibb, and T. Greene for useful
information and discussions about various sources in our sample.
KEL acknowledges K. Eastwood 
and the Northern Arizona University Research Experience for
Undergraduates program through NSF grant AST-0453611.  These observations
benefitted from the expert telescope operating skills of B. Golisch.  
We are grateful for a detailed and timely anonymous referee report
which improved the presentation of this paper.  The authors
extend special thanks to those of Hawaiian ancestry on whose sacred
mountain we are privileged to be guests.
This work made use of the SIMBAD reference database, the NASA
Astrophysics Data System, and the data products from the Two Micron All
Sky Survey, which is a joint project of the University of Massachusetts
and the Infrared Processing and Analysis Center/California Institute
of Technology, funded by the National Aeronautics and Space
Administration and the National Science Foundation.

{}

\clearpage
\begin{deluxetable}{llll}
\tablewidth{0pt}
\tablecaption{Sample\label{tbl-1}}
\tablehead{
\colhead{Object} & \colhead{$\alpha$} & \colhead{$\delta$} & \colhead{UT Date of}\\
\colhead{Name} & \colhead{(J2000.0)} & \colhead{(J2000.0)} & \colhead{Observation}}
\startdata
IRAS 04016+2610 & 04 04 43.0 & +26 18 57 & 2000 Nov 18\\
IRAS 04181+2654A & 04 21 11.5 & +27 01 09 & 2000 Nov 18\\
IRAS 04239+2436\tablenotemark{a} & 04 26 57.1 & +24 43 36 & 2000 Nov 18\\
Haro 6-10\tablenotemark{b} & 04 29 24.2 & +24 33 00 & 2000 Nov 19\\
L1551 IRS 5\tablenotemark{c} & 04 31 34.0 & +18 08 05 & 2000 Nov 18\\
IRAS 04361+2547\tablenotemark{d} & 04 39 13.9 & +25 53 21 & 2000 Nov 19\\
IRAS 04489+3042 & 04 52 06.7 & +30 47 18 & 2000 Nov 20
\enddata
\tablenotetext{a}{Binary of separation 0\farcs30 \citep{rei00}.}
\tablenotetext{b}{Binary of separation 1\farcs2 \citep{lei89}; listed as
Haro 6-10~S (visible light primary) and N in subsequent tables.  Haro 6-10~S
is a radial velocity variable and thus possibly a spectroscopic
binary \citep{dop08}.}
\tablenotetext{c}{Binary of separation 0\farcs35 \citep{loo97}.}
\tablenotetext{d}{Binary of separation 0\farcs31 \citep{ter98}.}
\end{deluxetable}

\clearpage
\begin{deluxetable}{lccl}
\tablewidth{0pt}
\tablecaption{Spectral Type Standards\label{tbl-2}}
\tablehead{
\colhead{Object} & \colhead{Spectral} & \colhead{$K_s$} & \colhead{UT Date of}\\
\colhead{Name} & \colhead{Type} & \colhead{(mag)} & \colhead{Observation}}
\startdata
HR 996 & G5 V & 3.0 & 2000 Nov 19\\
HR 995 & G6 IV & 4.1 & 2000 Nov 20\\ 
HR 7957 & K0 IV & 1.4 & 2000 Nov 20\\
HR 166 & K0 V & 4.0 & 2000 Nov 19\\
HR 753 & K3 V & 3.5 & 2000 Nov 19\\
GL 846 & M0.5 V & 5.3 & 2000 Nov 18\\
GL 908 & M1 V & 5.0 & 2000 Nov 18\\
GL 15A & M1.5 V & 4.0 & 2000 Nov 20\\
GL 806 & M2 V & 6.5 & 2000 Nov 19\\
GL 752A & M3 V & 4.7 & 2000 Nov 18\\
HR 9064 & M3 III & -0.1 & 2000 Nov 20\\
GL 876 & M4 V & 5.0 & 2000 Nov 20\\
GL 83.1 & M4.5 V & 6.6 & 2000 Nov 18\\
\enddata
\end{deluxetable}

\clearpage
\begin{deluxetable}{lcccccc}
\rotate
\tablewidth{0pt}
\tablecaption{Sample Photometry\label{tbl-3}}
\tablehead{
\colhead{Object} & \colhead{$J$} & \colhead{$H$} & \colhead{$K_s$} & \colhead{$L$ }  & \colhead{$L$-band} & \colhead{Color-Color} \\
\colhead{Name} & \colhead{(mag)} & \colhead{(mag)} & \colhead{(mag)} & \colhead{(mag)} & \colhead{References} & \colhead{Diagram $A_v$ (mag)} }
\startdata
IRAS 04016+2610 & 14.01$\pm$0.07 & 12.16$\pm$0.08 & 9.84\tablenotemark{a} & 6.8\tablenotemark{a} & 1 & 0 $\pm$1.6\\
IRAS 04181+2654A & 16.22$\pm$0.08 & 12.65$\pm$0.02 & 10.34$\pm$0.03 & 8.5\tablenotemark{b} & 2 & 22.8 $\pm$1.2\\
IRAS 04239+2436 & 15.75$\pm$0.09 & 12.35$\pm$0.04 & 9.99$\pm$0.02 & 7.2\tablenotemark{b} & 2 & 20.0 $\pm$1.4\\
Haro 6-10~S\tablenotemark{c} & 11.54$\pm$0.03\tablenotemark{d} & 10.6$\pm$0.1 & 8.6$\pm$0.1\tablenotemark{a} & 6.1$\pm$0.1\tablenotemark{b} & 3 & 0 $\pm$1.9\\
Haro 6-10~N\tablenotemark{c} & $...$ & 13.5$\pm$0.1 & 8.7$\pm$0.1\tablenotemark{a} & 4.9$\pm$0.1\tablenotemark{b} & 3 & $...$\\
L1551 IRS 5 & 13.71$\pm$0.06 & 11.51$\pm$0.05 & 9.82$\pm$0.04 & 7.4\tablenotemark{a} & 4 & 9.1 $\pm$ 1.2\\
IRAS 04361+2547 & 16.44$\pm$0.12 & 13.02$\pm$0.04 & 10.72$\pm$0.03 & 8.9\tablenotemark{a} & 5 & 20.8 $\pm$ 1.8\\
IRAS 04489+3042 & 14.43$\pm$0.03 & 12.02$\pm$0.02 & 10.38$\pm$0.02 & 1.5\tablenotemark{a} & 6 & 12.3 $\pm$ 0.6\\
\enddata
\tablecomments{Data are from 2MASS unless otherwise specified.}
\tablenotetext{a}{No uncertainty available.}
\tablenotetext{b}{$L'$ magnitude.}
\tablenotetext{c}{$H$, $K$, and $L'$ magnitudes quoted for the highly variable Haro 6-10
system are from reference (3) and correspond to the date of their observations
closest in time to our spectroscopy.}
\tablenotetext{d}{All $J$-band flux from Haro 6-10 is assumed to come from the ``primary'', southern
component; the J magnitude is from 2MASS and was not observed concurrently
with the H, K, and L data.}
\tablerefs{(1) Benson et al. (1984), (2) Beck (2007), (3) Leinert et al. (2001),
(4) Cohen \& Schwartz (1983),
(5) Kenyon et al. (1990), (6) Myers et al. (1987)}
\end{deluxetable}

\clearpage
\begin{deluxetable}{llcccccc}
\rotate
\tablewidth{0pt}
\tablecaption{Standard Star $K$-band Equivalent Widths (\AA)\label 
{tbl-4}}
\tablehead{
\colhead{$  $} & \colhead{Spectral} & \colhead{2.166$\mu$m}  & \colhead{2.208$\mu$m}  & \colhead{2.264$\mu$m}  & \colhead{2.281$\mu$m}  & \colhead{2.294$\mu$m} & \colhead{2.323$\mu$m} \\
\colhead{Name} & \colhead{Type} & \colhead{H~I} & \colhead{Na~I} & \colhead{Ca~I} &  \colhead{Mg~I} & \colhead{CO(2$-$0)} & \colhead{CO(3$-$1)}}
\startdata
HR 996 & G5 V  & 3.2 & 1.3 & 1.2 & 1.1   & 1.5 & 2.1 \\
HR 995 & G6 IV & 2.2 & 0.9 & 1.1 & 0.7   & 4.7 & 5.0 \\
HR 7957 & K0 IV& 2.1 & 0.9 & 1.4 & $...$ & 5.9 & 7.1 \\
HR 166 & K0 V  & 2.1 & 1.5 & 1.6 & 1.7   & 4.5 & 6.5 \\
HR 753 & K3 V  & 1.4 & 2.2 & 2.1 & 1.9   & 6.6 & 6.6 \\
GL 846 & M0.5 V& $...$& 4.4 & 4.3 & 1.1 & 7.4 & 6.2 \\
GL 908 & M1 V & $...$ & 1.9 & 2.7 &$...$& 4.8 & 5.1 \\
GJ 15A & M1.5 V& $...$& 3.3 & 3.2 &$...$& 5.6 & 5.3 \\
GL 806 & M2 V & $...$ & 3.8 & 3.9 &$...$& 6.6 & 6.3 \\
GL 752A & M3 V &$...$ & 4.6 & 4.5 &$...$& 7.4 & 7.1 \\
HR 9064 & M3 III&$...$& 2.2 & 2.8 & 2.0  & 20.9 & 13.7 \\
GJ 876 & M4 V & $...$ & 6.2 & 4.2 & $...$ & 4.5 & 6.2 \\
GL 83.1 & M4.5 V&$...$& 5.1 & 2.4 & $...$  & 6.0 & 4.3
\enddata
\end{deluxetable}

\clearpage
\begin{deluxetable}{lccccccccc}
\rotate
\tabletypesize{\tiny}
\tablewidth{0pt}
\tablecaption{Target Star Equivalent Widths (\AA)\label 
{tbl-5}}
\tablehead{
\colhead{$  $}  & \colhead{2.059$\mu$m} & \colhead{2.122$\mu$m} & \colhead{2.166$\mu$m} & \colhead{2.208$\mu$m} & \colhead{2.264$\mu$m}  &  \colhead{2.281$\mu$m} & \colhead{2.294$\mu$m} & \colhead{2.323$\mu$m} & \colhead {4.052$\mu$m}\\
\colhead{Name}  & \colhead{He~I} & \colhead{H$_{2}$}  & \colhead{H~I} & \colhead{Na~I} & \colhead{Ca~I} &  \colhead{Mg~I} & \colhead{CO(2$-$0)} & \colhead{CO(3$-$1)} & \colhead{Br $\alpha$}}
\startdata
IRAS 04016$+$2610 & $...$ & $-$2.2 & $-$2.3 & 1.0  &  1.0 & 2.0  & 2.3 & $...$ & $-$11.5\\
IRAS 04181$+$2654A &$-$1.4 & $-$1.9 & $-$9.9 & 1.3 &  1.6 & $-$0.8 & $...$ &  4.0 & $-$23.0\\
IRAS 04239$+$2436 & $-$2.6 & $-$2.3 & $-$15.3 &$-$2.4 & 1.4 & $...$ & $-$11.9 & $-$10.5 & $-$25.0\\
Haro 6-10~S       & $...$ & $-$2.1 & $-$4.1 &  $...$ & 1.0 &  1.0 & $...$ &  $...$ & $-$33.0\\
Haro 6-10~N       & $...$ & $-$2.5 & $-$3.1 & $-$1.1 &$...$ & $...$ & $-$6.7 & $-$2.2 & $...$\\
L1551 IRS 5      & $...$ & $-$2.5 & $-$1.7 &  2.4 &  1.9 & $...$ &  16.7 &  14.8 & $-$4.5\\
IRAS 04361$+$2547 & $...$ & $-$2.8 & $-$2.6 & $...$ & 0.5 &  1.4 & $...$ & $...$ & $-$2.0\\
IRAS 04489$+$3042 & $-$2.6 & $...$  & $-$2.7 &  2.0 &  1.7 & $...$ &  4.6 &  4.5 & $...$
\enddata
\end{deluxetable}

\clearpage
\begin{deluxetable}{lcccc}
\tablewidth{0pt}
\tablecaption{Target Star Results\label 
{tbl-6}}
\tablehead{
\colhead{$  $} & \colhead{Spectral}  & \colhead{$  $} & \colhead{$  $} & \colhead{Log(L$_{acc}$)}\\
\colhead{Name} & \colhead{Type} & \colhead{$A_{v}$} & \colhead{$r_{k}$} & \colhead{L$_{\odot}$}}
\startdata
IRAS 04016+2610 & K3 V $\pm$ 2\tablenotemark{a} & 45 $\pm$ 2 & 1.3 $\pm$ 0.2 & 0.54 $\pm$ 0.99\\
IRAS 04181+2654A & M1 $-$ M4 & $<$34 & $...$ & 0.35 $\pm$ 1.04\\
IRAS 04239+2436 & G $-$ early K? & $<$45  & $...$ & 1.36 $\pm$ 0.95\\
Haro 6-10~S & early K & 30 $\pm$ 2 & 0.5 $\pm$ 0.1 & 0.37 $\pm$ 1.04\\
Haro 6-10~N & G $-$ early K? & $<$59 & $...$ & 2.06 $\pm$ 0.91\\
L1551 IRS 5 & M3 III $\pm$1 & 28 $\pm$ 2 & 1.0 $\pm$ 0.2 & $-$0.69 $\pm$ 1.12\\
IRAS 04361+2547 & M4 V $\pm$ 2 & $<$30 & $...$ & $-$0.80 $\pm$ 1.13\\
IRAS 04489+3042 & M3.5 IV $\pm$ 2 & 18 $\pm$ 1 & 1.4 $\pm$ 0.4 & $-$1.26 $\pm$ 1.17
\enddata
\tablenotetext{a}{Uncertainties in spectral type refer to number of
spectral subclasses.}
\end{deluxetable}

\clearpage
\begin{deluxetable}{lcc}
\tablewidth{0pt}
\tablecaption{Ice Feature Optical Depths\label{tbl-7}}
\tablehead{
\colhead{Name} & \colhead{$\tau_{ice}$} & \colhead{$A_{v}$}}
\startdata
IRAS 04016+2610 & 2.8 & 31.7 \\
IRAS 04181+2654A & 1.0 & 12.4  \\
IRAS 04239+2436 & 1.4 & 16.7  \\
Haro 6$-$10 S & 0.7 & 9.1  \\
Haro 6$-$10 N & 0.5 & 7.0  \\
L1551 IRS 5 & 1.7 & 19.9  \\
IRAS 04361+2547 & 2.1 & 24.2  \\
IRAS 04489+3042 & 0.7 & 9.1 
\enddata
\tablecomments{Uncertainties in $\tau_{ice}$ are $\sim$0.1 and in $A_v$ are 1.1 mag.}
\end{deluxetable}

\clearpage
\begin{deluxetable}{lcccccc}
\rotate
\tabletypesize{\tiny}
\tablewidth{0pt}
\tablecaption{Input and Results from H-R Diagram\label 
{tbl-8}}
\tablehead{
\colhead{$  $} & \colhead{T$_{eff}$}  & \colhead{L$_*$ (L$_{\odot}$)} & \colhead{L$_*$ (L$_{\odot}$)} & \colhead{L$_*$ (L$_{\odot}$)} & \colhead{Age\tablenotemark{a}} & \colhead{Mass\tablenotemark{b}}\\
\colhead{Name} & \colhead{(K)} & \colhead{\citet{fur08}} & \colhead{\citet{whi04}} & \colhead{\citet{dop05}} & \colhead{(years)} & \colhead{(M$_{\odot}$)}}
\startdata
IRAS 04016+2610 & 4800 $\pm$ 200 & 0.45 $\pm$ 0.23 & 0.45 & 4.9 & 1e6\tablenotemark{c} & 2.0\tablenotemark{c}\\
IRAS 04181+2654A & 3350 $\pm$ 100 & 0.14 $\pm$ 0.04 & $...$ & $...$ & 2 & 0.2 \\
IRAS 04239+2436 & $...$ & 0.15 $\pm$ 0.51 & $...$ & $...$ & $...$ & $...$\\
Haro 6-10~S & 4800 $\pm$ 200 & $...$ & 1.8 & 3.3 & 2e6\tablenotemark{c} & 1.8\tablenotemark{c}\\
Haro 6-10~N & $...$ & $...$ & $...$ & $...$ & $...$ & $...$\\
L1551 IRS 5 & 3300 $\pm$ 200 & 7.50 $\pm$ 1.46 & $...$ & 2.6 & $<$1e6 & 0.2\\
IRAS 04361+2547 & 3200 $\pm$ 200 & 3.60 $\pm$ 0.41 & $...$ & $...$ & $<$1e6 & 0.2\\
IRAS 04489+3042 & 3250 $\pm$ 200 & 0.21 $\pm$ 0.03 & 0.2 & 0.3 & 1e6 & 0.2
\enddata
\tablenotetext{a}{Age uncertainties are $\sim$1~Myr.}
\tablenotetext{b}{Mass uncertainties are 0.1$-$0.2~M$_{\odot}$.}
\tablenotetext{c}{Based on luminosities from \citet{dop05}; all other results
based on luminosities from \citet{fur08}.}
\end{deluxetable}

\clearpage
\begin{deluxetable}{lccccc}
\rotate
\tabletypesize{\tiny}
\tablewidth{0pt}
\tablecaption{Accretion Luminosity and Mass Accretion Rate\label 
{tbl-9}}
\tablehead{
\colhead{$  $} & \colhead{log(L$_{acc}$/L$_{\odot}$)} & \colhead{log(L$_{acc}$/L$_{\odot}$)} & \colhead{log(L$_{acc}$/L$_{\odot}$)} & \colhead{$\dot{M}$ (M$_{\odot}$)} & \colhead{$\dot{M}$ (M$_{\odot}$)}\\
\colhead{Name} & \colhead{this work} & \colhead{\citet{muz98}} & \colhead{\citet{bec07}} & \colhead{this work} & \colhead{other}}
\startdata
IRAS 04016+2610 & 0.54 $\pm$ 0.99 & $-$1.52 & $...$ & 2.2e$-$7 & 7.1e$-$8\tablenotemark{a}\\
IRAS 04181+2654A & 0.35 $\pm$ 1.04 & $...$ & $-$1.22 & 4.9e$-$7 & 9.0e$-$9\tablenotemark{b}\\
IRAS 04239+2436 & 1.36 $\pm$ 0.95 & $-$0.62 & $-$0.15 & $...$ & $...$\\
Haro 6-10~S & 0.37 $\pm$ 1.04 & $...$ & $...$ & 1.4e$-$7 & 2.0e$-$7\tablenotemark{a}\\
Haro 6-10~N & 2.06 $\pm$ 0.91 & $-$1.34 & $...$ & $...$ & $...$\\
L1551 IRS 5 & $-$0.69 $\pm$ 1.12 & $...$ & $...$ & 3.4e$-$7 & $...$\\
IRAS 04361+2547 & $-$0.80 $\pm$ 1.13 & $-$1.10 & $-$0.70 & 1.9e$-$7 & $...$\\
IRAS 04489+3042 & $-$1.26 $\pm$ 1.17 & $-$1.70 & $...$ & 1.6e$-$8 & 4.1e$-$10\tablenotemark{a}\\
\enddata
\tablenotetext{a}{\citet{whi04}.}
\tablenotetext{b}{\citet{bec07}.}
\end{deluxetable}

\clearpage
\begin{figure}
\plotone{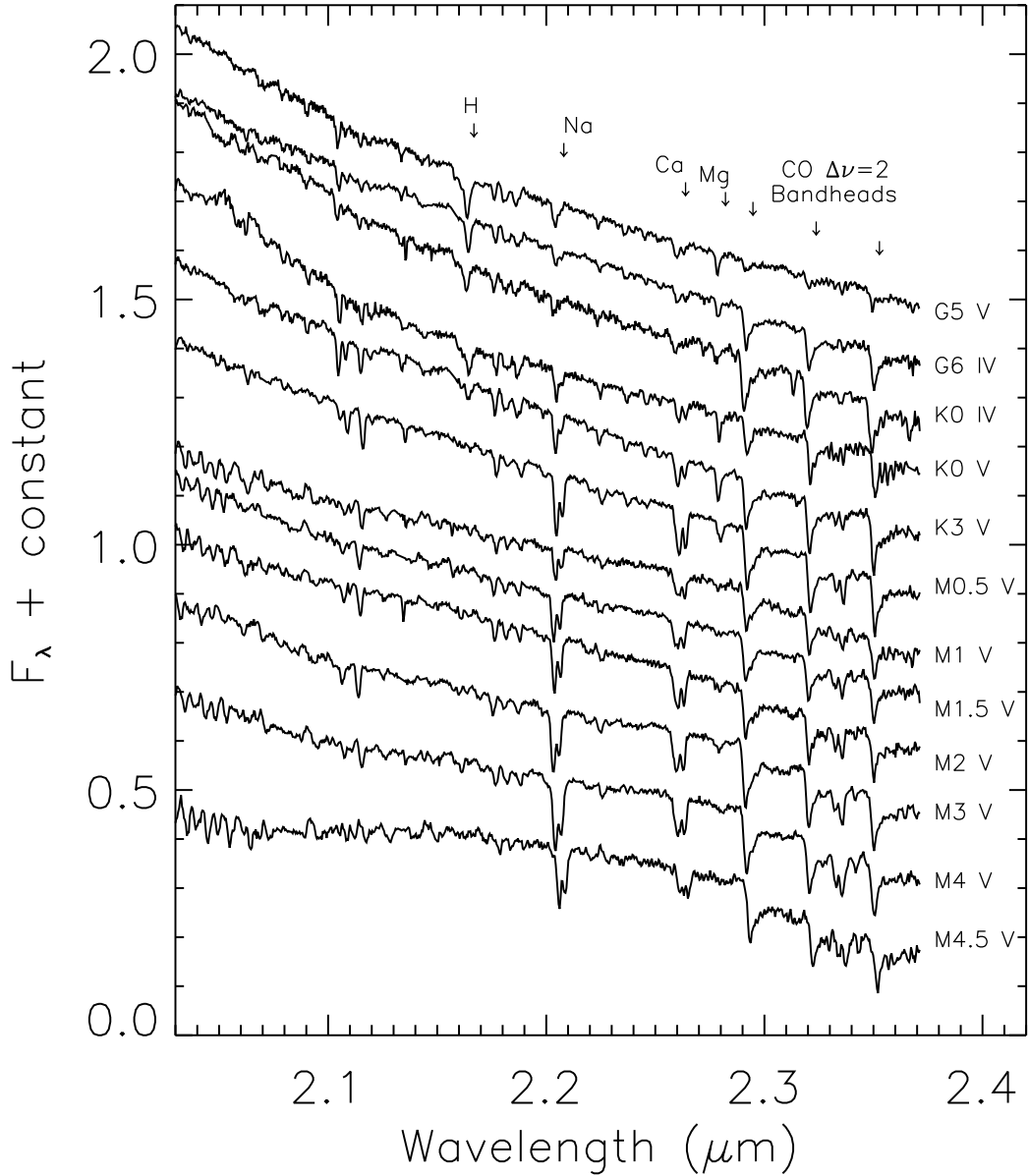}
\caption{$K$-band spectra of the dwarf and subgiant spectral type standards.
The stars are identified in Table 2. Prominent spectral features are indicated.
The spectra were normalized
to unity at 2.2~$\mu$m and shifted by a constant for presentation.
\label{fig1}}
\end{figure}

\clearpage
\begin{figure}
\plotone{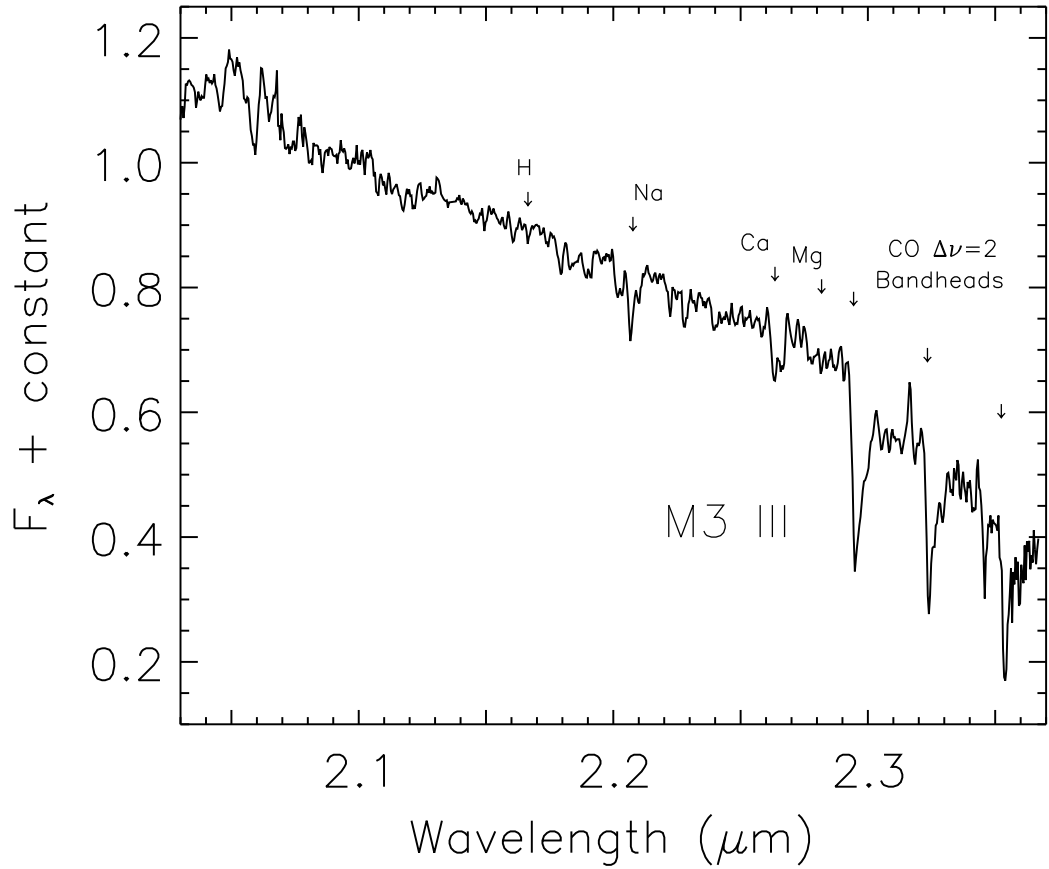}
\caption{$K$-band spectrum of the giant spectral type standard. The star
is identified in Table 2. Prominent spectral features are indicated.  The
spectrum was normalized
to unity at 2.2~$\mu$m and shifted by a constant for presentation.
\label{fig2}}
\end{figure}

\clearpage
\begin{figure}
\plotone{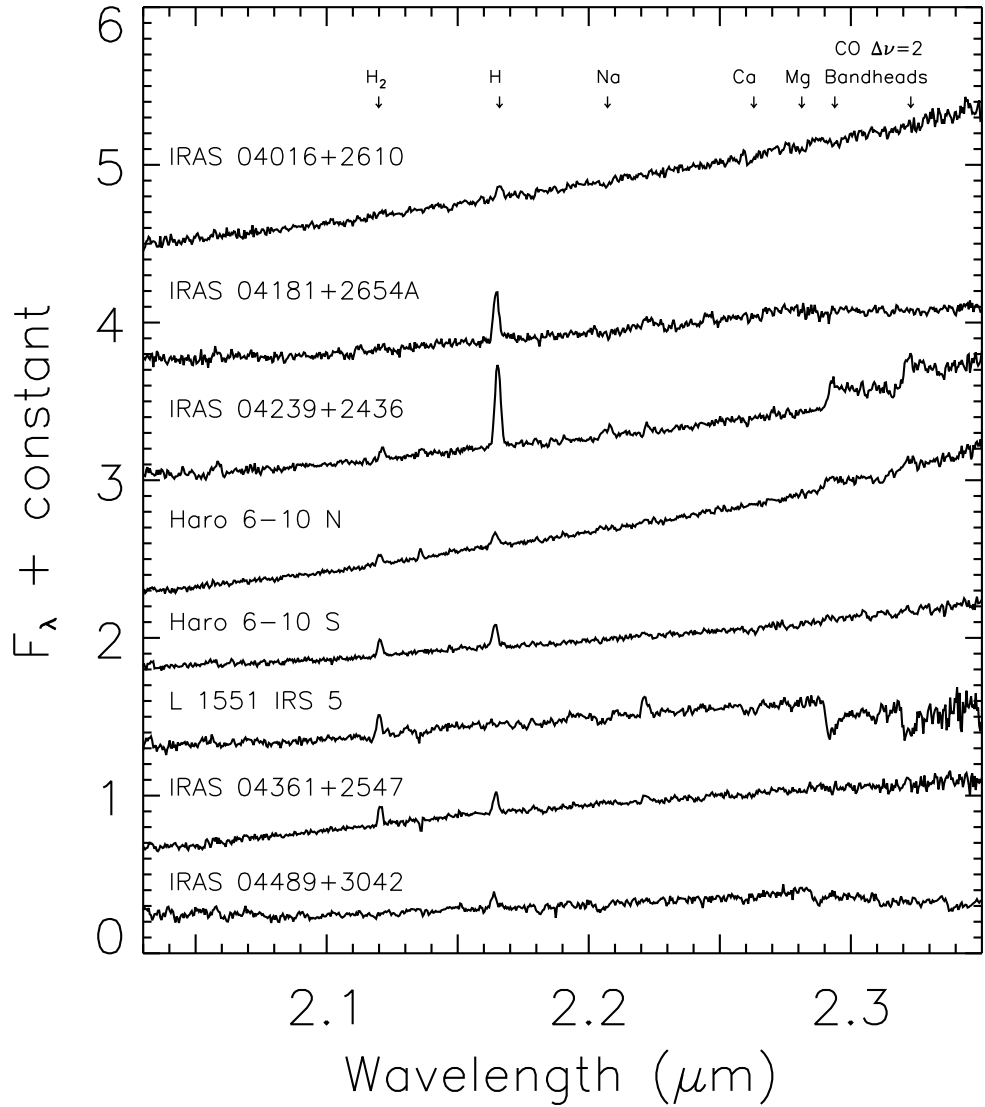}
\caption{$K$-band spectra for our sample of
Class I sources. Prominent spectral
features are indicated.  The spectra were normalized
to unity at 2.2~$\mu$m and shifted by a constant for presentation.
\label{fig3}}
\end{figure}

\clearpage
\begin{figure}
\plotone{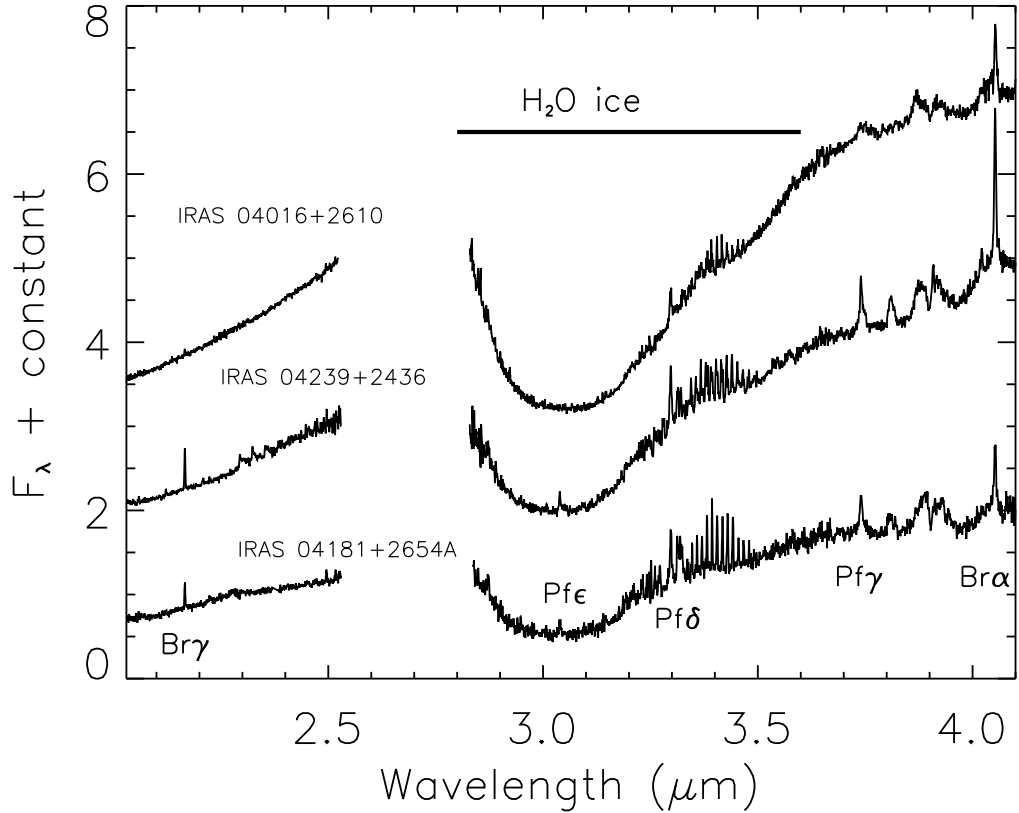}
\caption{$K$- and $L$-band spectra for the spectroscopic sample;
the region around 2.7~$\mu$m is contaminated by terrestrial water absorption
is not shown.  The spectra were extracted from five overlapping orders and
scaled to match in the overlap regions.
The overall spectrum was normalized at 2.25~$\mu$m and
shifted by a constant for presentation. The ice band
feature and hydrogen lines are indicated.  Pfund series emission lines
are contaminated in part by residual inherent A0 star absorption lines
(Brackett lines are not contaminated).
\label{fig4a}}
\end{figure}

\clearpage
\setcounter{figure}{3}
\begin{figure}
\plotone{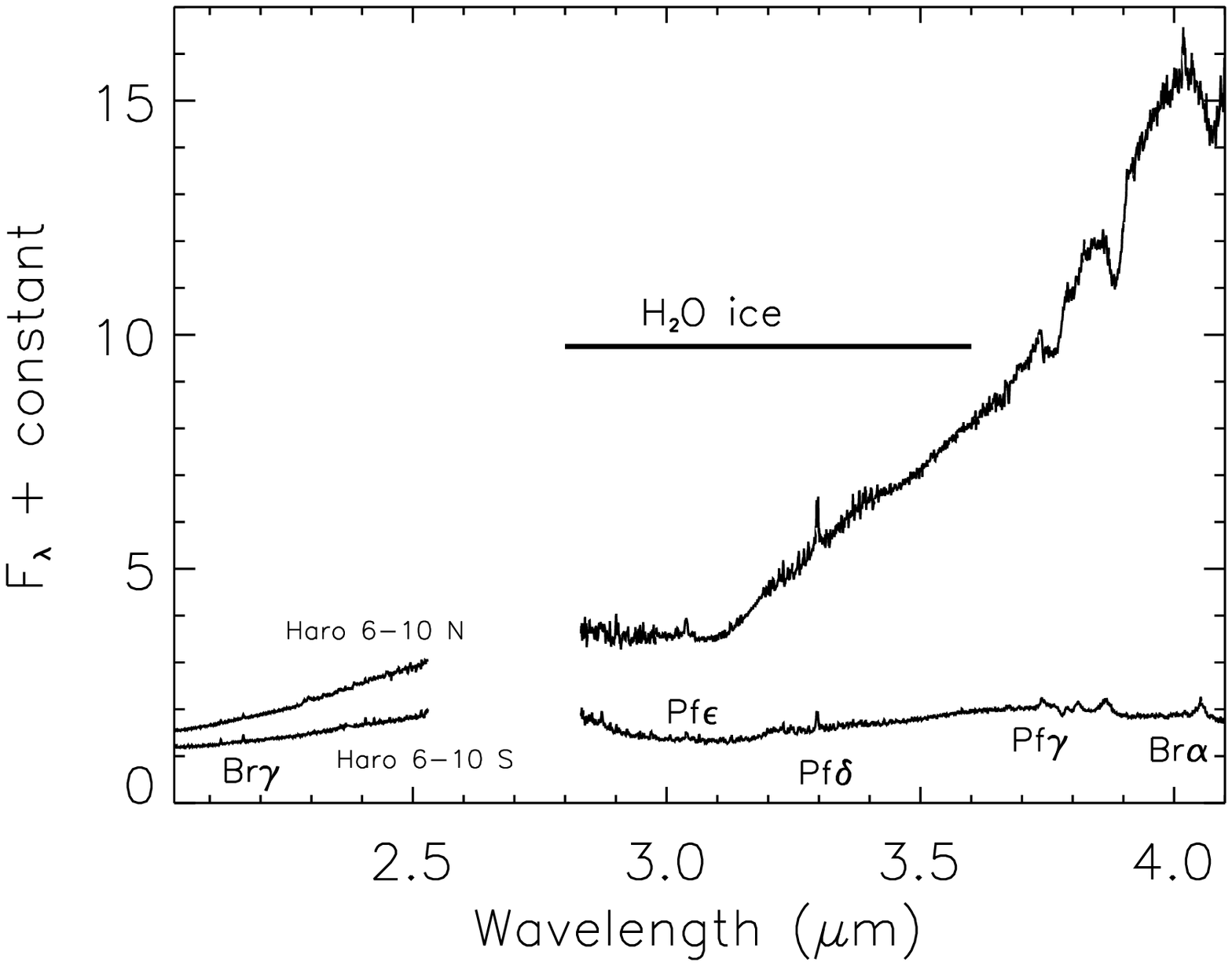}
\caption{Con'd.
\label{fig4b}}
\end{figure}

\clearpage
\setcounter{figure}{3}
\begin{figure}
\plotone{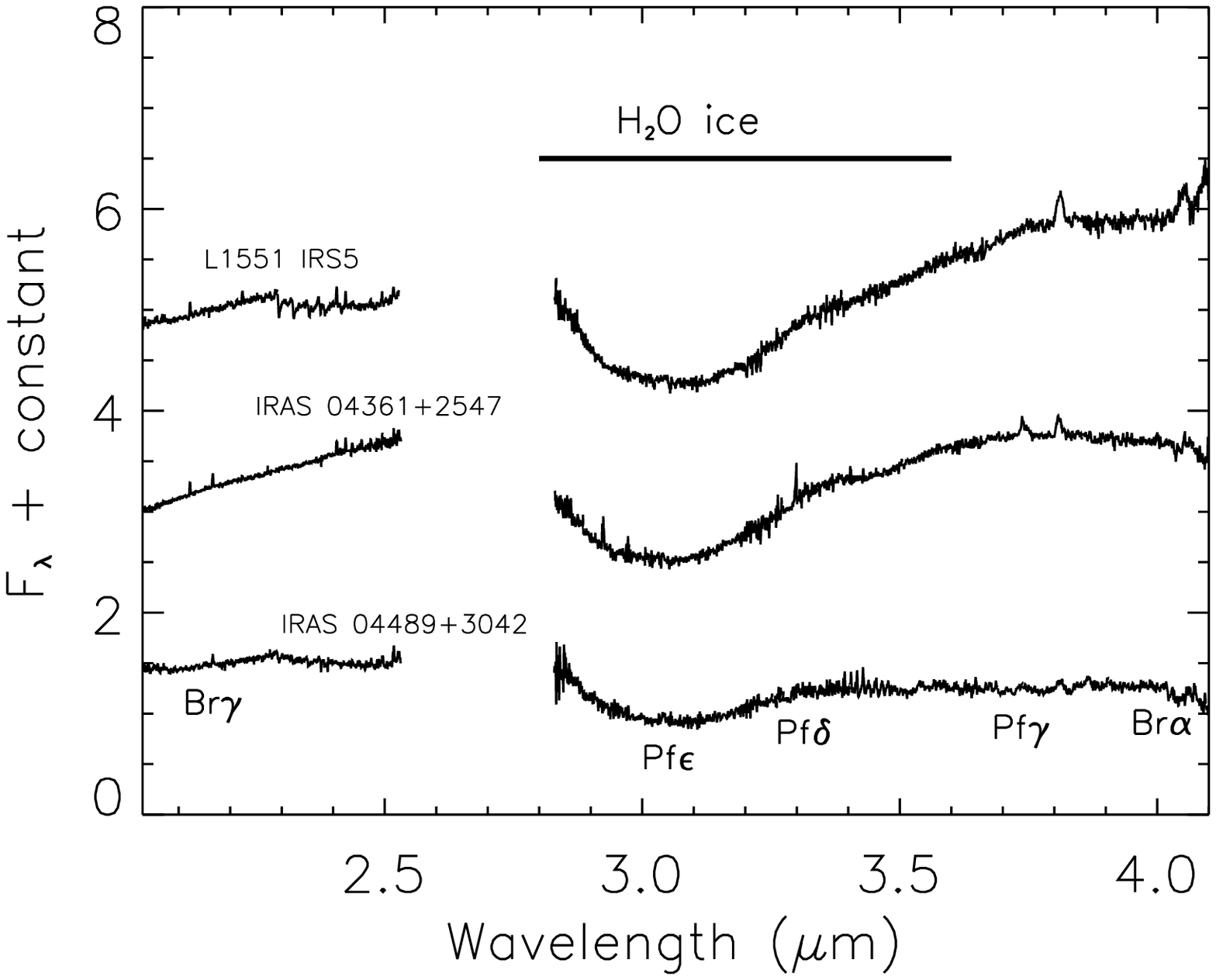}
\caption{Con'd.
\label{fig4c}}
\end{figure}

\clearpage
\begin{figure}
\plotone{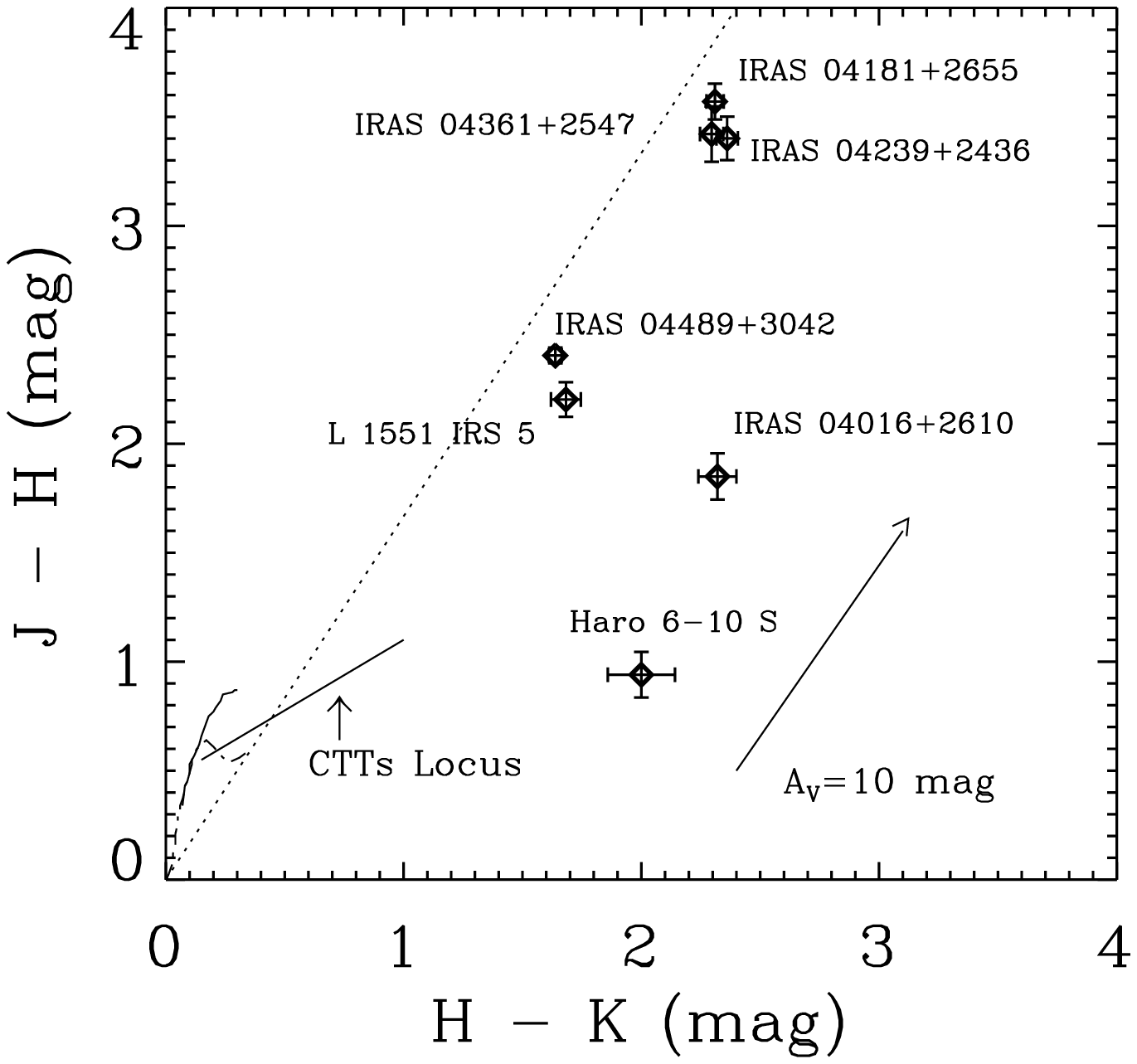}
\caption{J-H, H-K color-color diagram of the Class I sample,
except for Haro 6-10~N. Magnitude data were
obtained from 2MASS and from the literature. The dotted line separates
objects with (to the right and below) and without a near-IR excess.
The dwarf (dash-dot) and giant (solid) star loci are overplotted at the bottom left.
The CTTs locus and $A_{v}=10$ mag reddening vector from Meyer et al. (1997)
are also shown.
\label{fig5}}
\end{figure}

\clearpage
\begin{figure}
\plotone{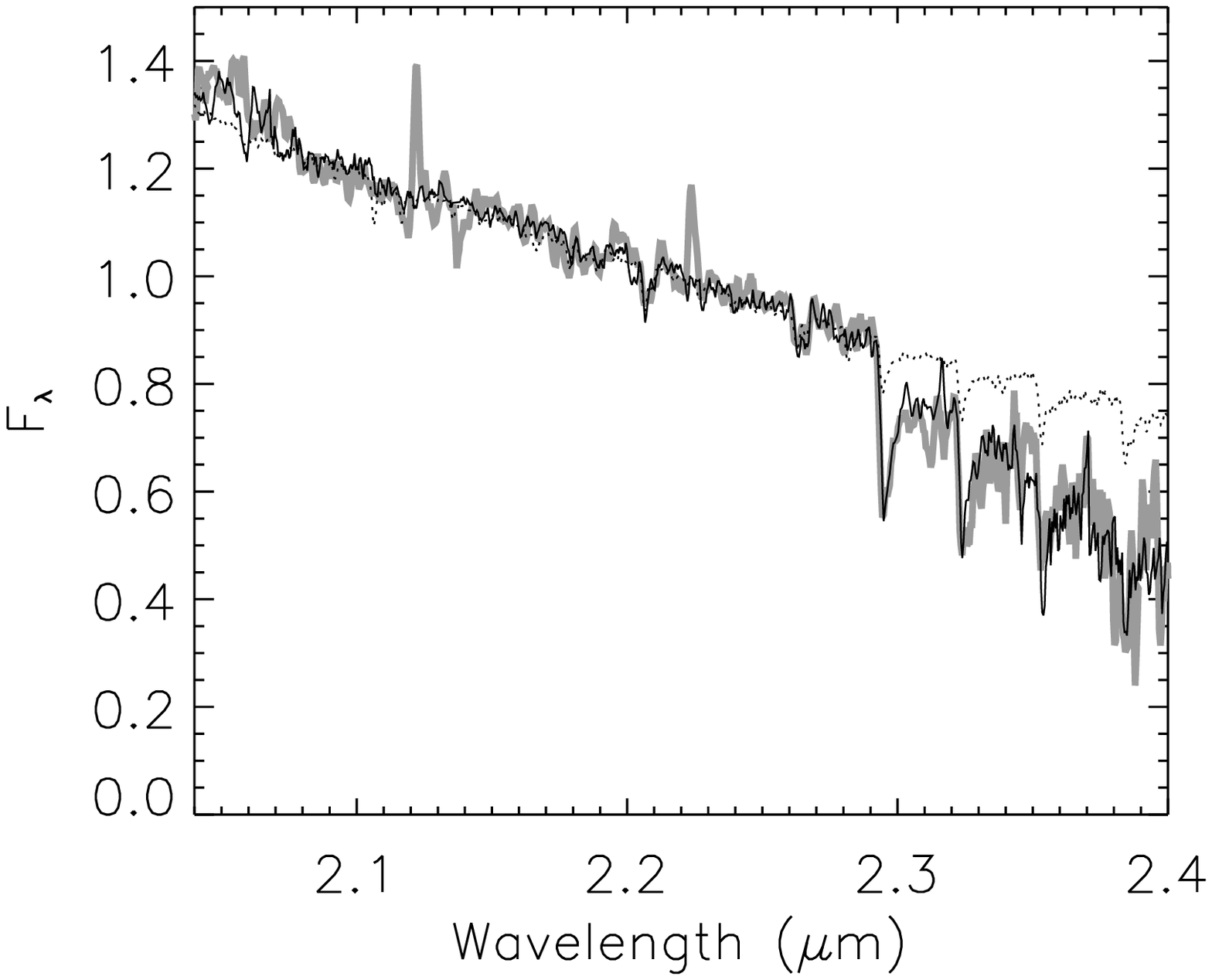}
\caption{The $K$-band spectrum of L1551 IRS 5 (grey), accounting for
28 magnitudes of visual extinction and r$_k=1.0$, overplotted with the best fit
stellar template (black), M3 III.  The dotted line shows a K3V standard
star, corresponding to the spectral type identified by \citet{dop05}.
\label{fig6}}
\end{figure}

\clearpage
\begin{figure}
\plotone{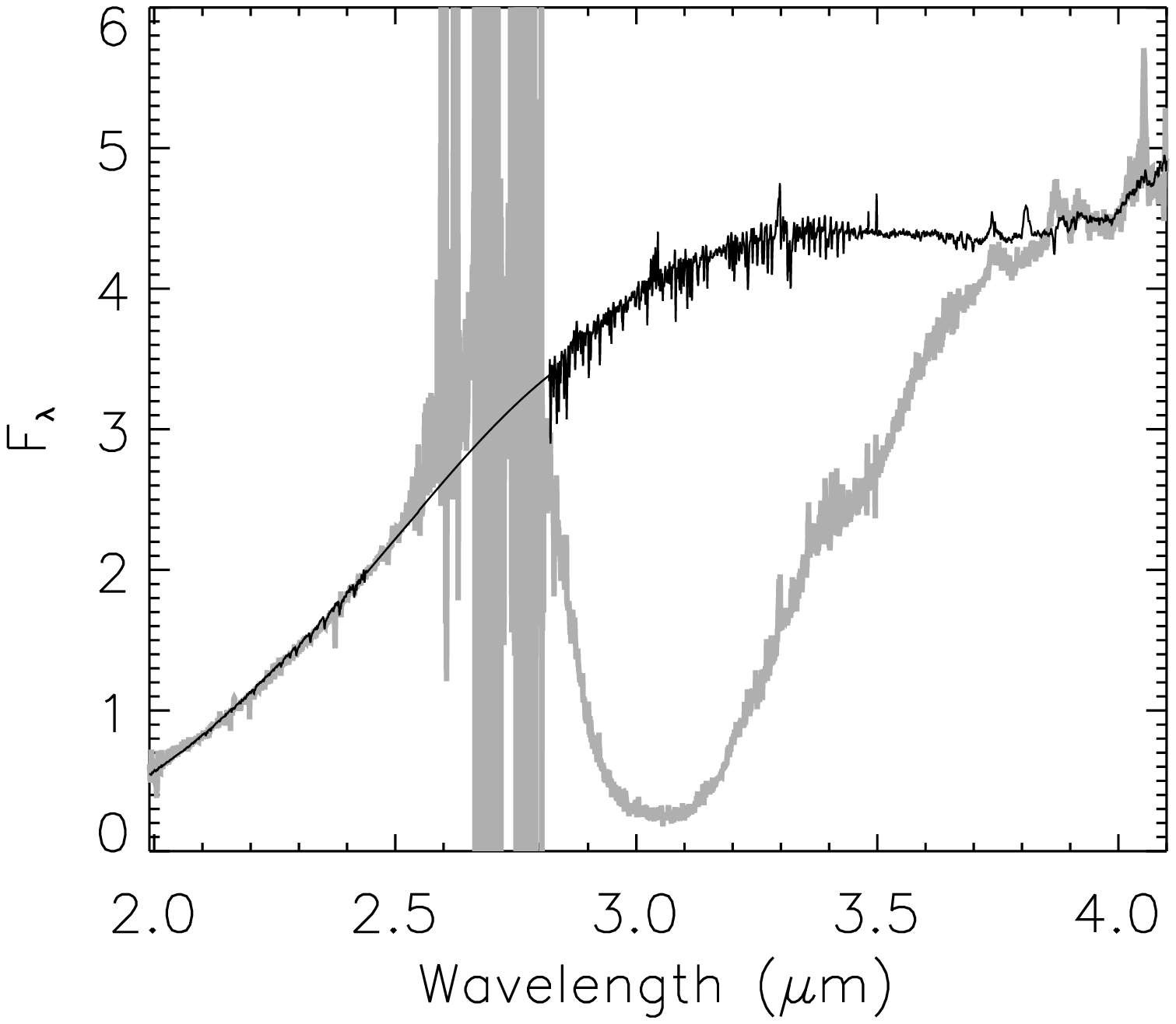}
\caption{The full $K$- and $L$-band spectrum of IRAS 04016$+$2610 (grey)
and the best fit modeled stellar template (black) used to determine the
optical depth of the 3.1~$\mu$m water ice feature.  The broad, noisy
section of spectrum from 2.6$-$2.8~$\mu$m corresponds to a region of
low transmission in the Earth's atmosphere.
\label{fig7}}
\end{figure}

\clearpage
\begin{figure}
\plotone{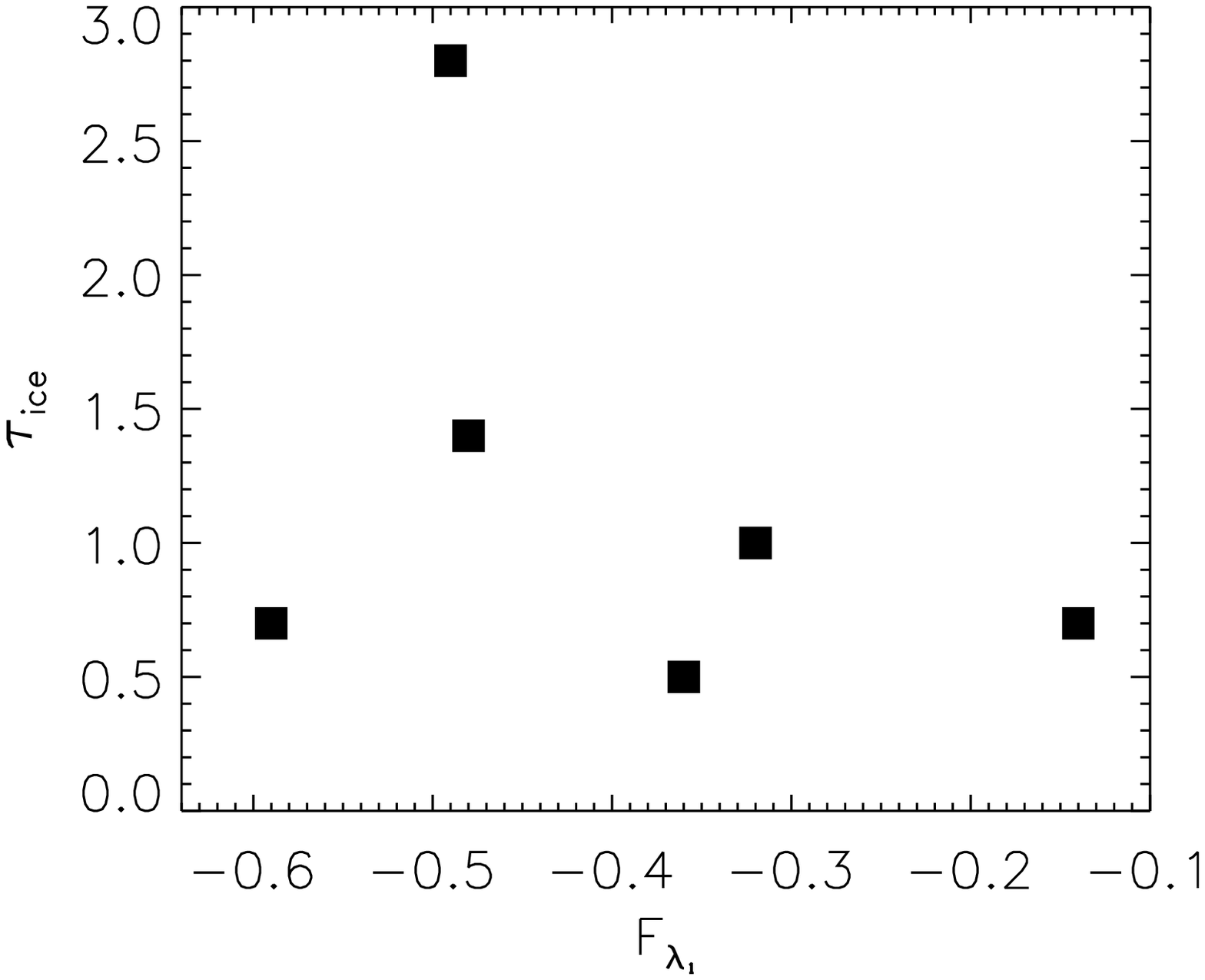}
\caption{The 3.1~$\mu$m ice optical depth compared to the strength of the
10~$\mu$m silicate absorption feature as measured by \citet{kes05}.
The data show a weak trend towards higher values of $\tau_{ice}$ among
objects with the strongest silicate features.
\label{fig8}}
\end{figure}

\clearpage
\begin{figure}
\plotone{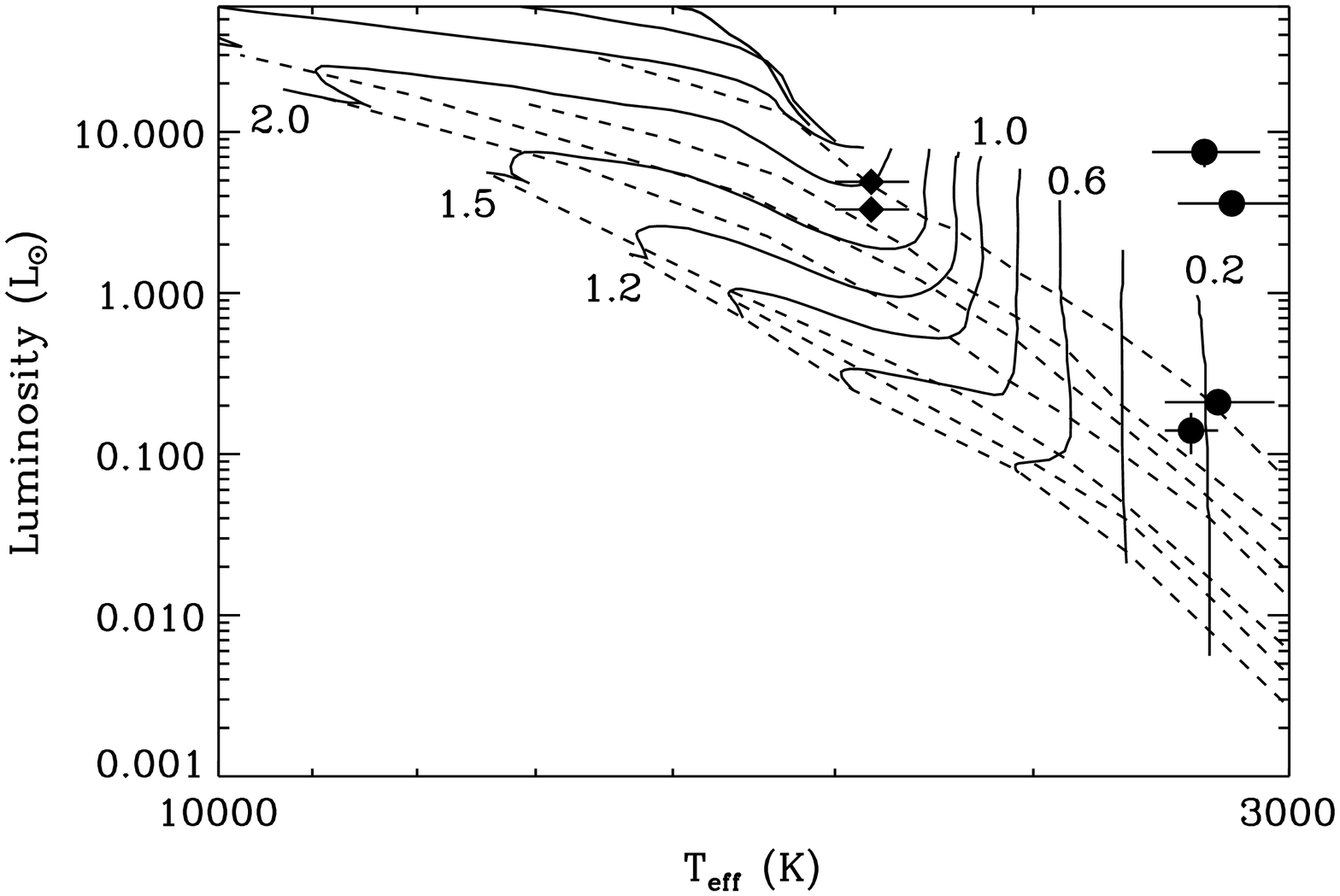}
\caption{Class I targets, for which we have estimated T$_{eff}$
and use values for the luminosity from \citet{fur08} (circles) and
\citet{dop05} (diamonds), plotted on
an H-R diagram with the evolutionary tracks of \citet{pal99}.
At the low-mass end, only
every other mass track is labeled.  Isochrones are plotted for
ages of 1e6, 3e6, 5e6, 1e7, 3e7, 5e7, and 1e8 years.  The two
systems in the upper right above the tracks are both unresolved subarcsecond
binaries; the other four that all fall on the tracks are considered to be single,
although \citet{dop08} suggests that Haro 6-10S is a spectroscopic binary.
\label{fig9}}
\end{figure}

\clearpage
\begin{figure}
\plotone{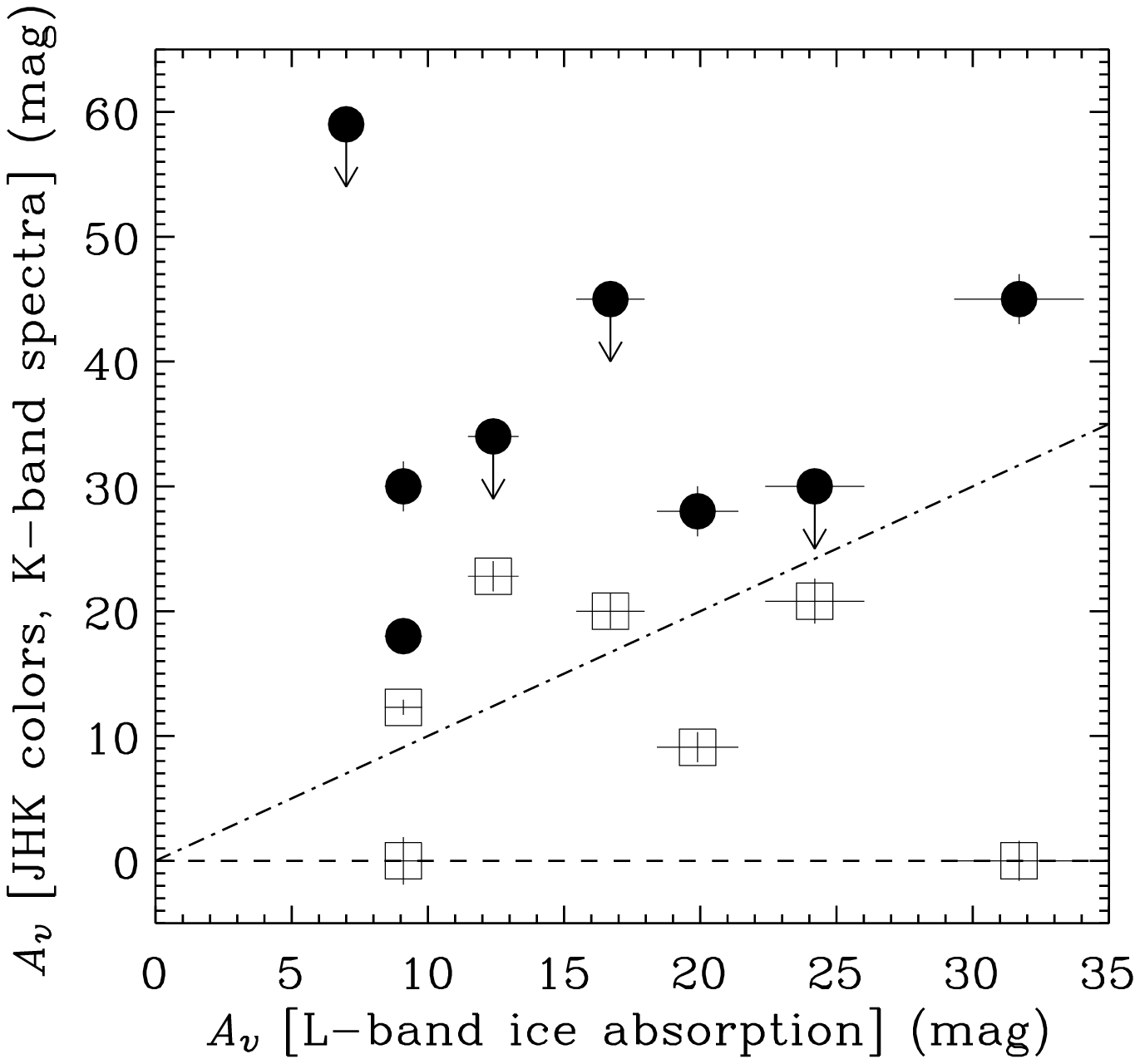}
\caption{Extinctions derived from the JHK color-color diagram
(open squares) and $K$-band spectral fitting (filled circles) as a function of
extinction derived from fitting the 3.1~$\mu$m ice feature.  The 
dash-dot line indicates equal extinctions on both axes.
\label{fig10}}
\end{figure}

\end{document}